\documentclass[pdflatex,sn-mathphys-num,iicol]{sn-jnl}

\usepackage{graphicx}%
\usepackage{multirow}%
\usepackage{amsmath,amssymb,amsfonts}%
\usepackage{amsthm}%
\usepackage{mathrsfs}%
\usepackage[title]{appendix}%
\usepackage{xcolor}%
\usepackage{textcomp}%
\usepackage{manyfoot}%
\usepackage{booktabs}%
\usepackage{algorithm}%
\usepackage{algorithmicx}%
\usepackage{algpseudocode}%
\usepackage{listings}%

\theoremstyle{thmstyleone}%
\theoremstyle{thmstyletwo}%

\theoremstyle{thmstylethree}%

\begin{document}

\title{A coupled geometric surface and surface reaction--diffusion framework for modelling nuclear mechanics during confined cell migration}


\author*[1]{\fnm{Francesca} \sur{Ballatore}}\email{francesca.ballatore@univ-cotedazur.fr}

\author[2]{\fnm{Silvia} \sur{Comunian}}\email{silviacomunian31@gmail.com}

\author[3,4,5,6]{\fnm{Anotida} \sur{Madzvamuse}}\email{am823@math.ubc.ca}

\author[2]{\fnm{Slimane} \sur{Ait-Si-Ali}}\email{slimane.aitsiali@u-pariscite.fr}

\author[1,7]{\fnm{Rachele} \sur{Allena}}\email{rachele.allena@univ-cotedazur.fr}
\affil*[1]{\orgname{Université Côte d'Azur, CNRS},  \orgdiv{Laboratoire Jean-Alexandre Dieudonné, UMR7351}, \orgaddress{\postcode{06108} \city{Nice}, \country{France}}}

\affil[2]{ \orgname{Université Paris Cité, CNRS}, \orgdiv{Epigenetics and Cell Fate, UMR7216}, \orgaddress{\postcode{F-75013} \city{Paris}, \country{France}}}

\affil[3]{\orgname{University of British Columbia}, \orgdiv{Mathematics Department}, \orgaddress{\postcode{V6T 1Z2} \city{Vancouver}, \country{Canada}}}

\affil[4]{\orgname{University of Zimbabwe},  \orgdiv{Department of Mathematics and Computational Sciences}, \orgaddress{\city{Mt Pleasant, Harare}, \country{Zimbabwe}}}

\affil[5]{\orgname{University of Pretoria},  \orgdiv{Department of Mathematics and Applied Mathematics}, \orgaddress{\postcode{0132} \city{Pretoria}, \country{South Africa}}}

\affil[6]{\orgname{University of Johannesburg}, \orgdiv{Applied Mathematics},  \orgaddress{\postcode{2006} \city{Johannesburg}, \country{South Africa}}}

\affil[7]{\orgname{Institut Universitaire de France}, \orgaddress{\postcode{75231} \city{Paris}, \country{France}}}

\abstract{Confined cell migration is governed by nuclear and cellular deformation. In this work, we formulate a two-dimensional mathematical framework in which the cell and nuclear membranes are represented as evolving closed surfaces governed by force-balance laws incorporating surface tension, bending elasticity, and external mechanical interactions. Nuclear mechanics depends on several factors, including chromatin organization and compaction, which are regulated by epigenetic modifications. Molecular regulation at the nuclear periphery is described by a surface reaction–diffusion system posed on the evolving nuclear surface. We apply this coupled framework to an experimentally characterized lung cancer cell model in which the balance between the enzymes SETDB1 and SUV39H1 regulates the peripheral deposition of H3K9me3. Using experimentally measured mechanical parameters, we represent control (Ctrl) and SETDB1 knockout (SETDB1 KO) conditions, which display distinct H3K9me3 distributions, especially at the nuclear periphery. Numerical solutions obtained using the evolving surface finite element method qualitatively reproduce the changes in nuclear mechanics as well as part of the experimentally observed increase in the entry time in SETDB1 KO cells. The model therefore reproduces the direction of the experimental observations, while the residual difference between simulated and measured entry times indicates that the cellular response is multifactorial.}

\keywords{cell translocation, confined environments, nuclear mechanics, epigenetic modifications, chromatin modification, geometric surface PDE (GS-PDE), surface reaction-diffusion system, evolving surface finite element method}



\maketitle

\section{Introduction}\label{sec1}
Migration under confinement underlies a wide range of physiological and pathological events, spanning immune surveillance \cite{Friedl:2008}, embryonic development \cite{Montell:2003, Uriu:2014}, and the metastatic spread of cancer\cite{Friedl:2003, Warner:2018}. Within dense interstitial tissues, cells frequently migrate through pre-existing pores that are smaller than their own diameter \cite{Wisdom:2018, Saraswathibhatla:2023, Wolf:2003}. To pass through these narrow spaces, both the cell and its nucleus must undergo substantial deformation. Being the largest and most rigid organelle within the cell, the nucleus typically constitutes the principal mechanical obstacle to migration through confined spaces \cite{Wolf:2013}. Microfluidic devices containing narrow constrictions provide a controlled \emph{in vitro} platform to study this process and quantify the time required for cells to deform and pass through confinement \cite{Guillou:2016, Chen:2020, Jebane:2023, AbarcaOrtega:2024}.

Nuclear deformability depends on several structural components, including chromatin, the complex of DNA and associated proteins that occupies most of the nuclear interior. DNA is wrapped around histone proteins and organized into either relatively accessible euchromatin or more compact heterochromatin. Heterochromatin accumulation near the nuclear envelope contributes to nuclear stiffness and resistance to deformation. These mechanical properties can be investigated using optical tweezers, which apply controlled forces to the nucleus and allow its deformation to be quantified \cite{Polimeno:2018}.

Chromatin organization is regulated by epigenetic modifications of DNA and histone proteins. One such modification is the trimethylation of histone H3 at lysine 9 (H3K9me3), a characteristic mark of heterochromatin that is enriched at the nuclear periphery \cite{Mozzetta:2015}. The histone methyltransferases SETDB1 and SUV39H1 contribute to H3K9me3 deposition and act in balance to regulate its spatial distribution within the nucleus.

\begin{figure*}[htbp]
\centering
\includegraphics[width=0.6\linewidth]{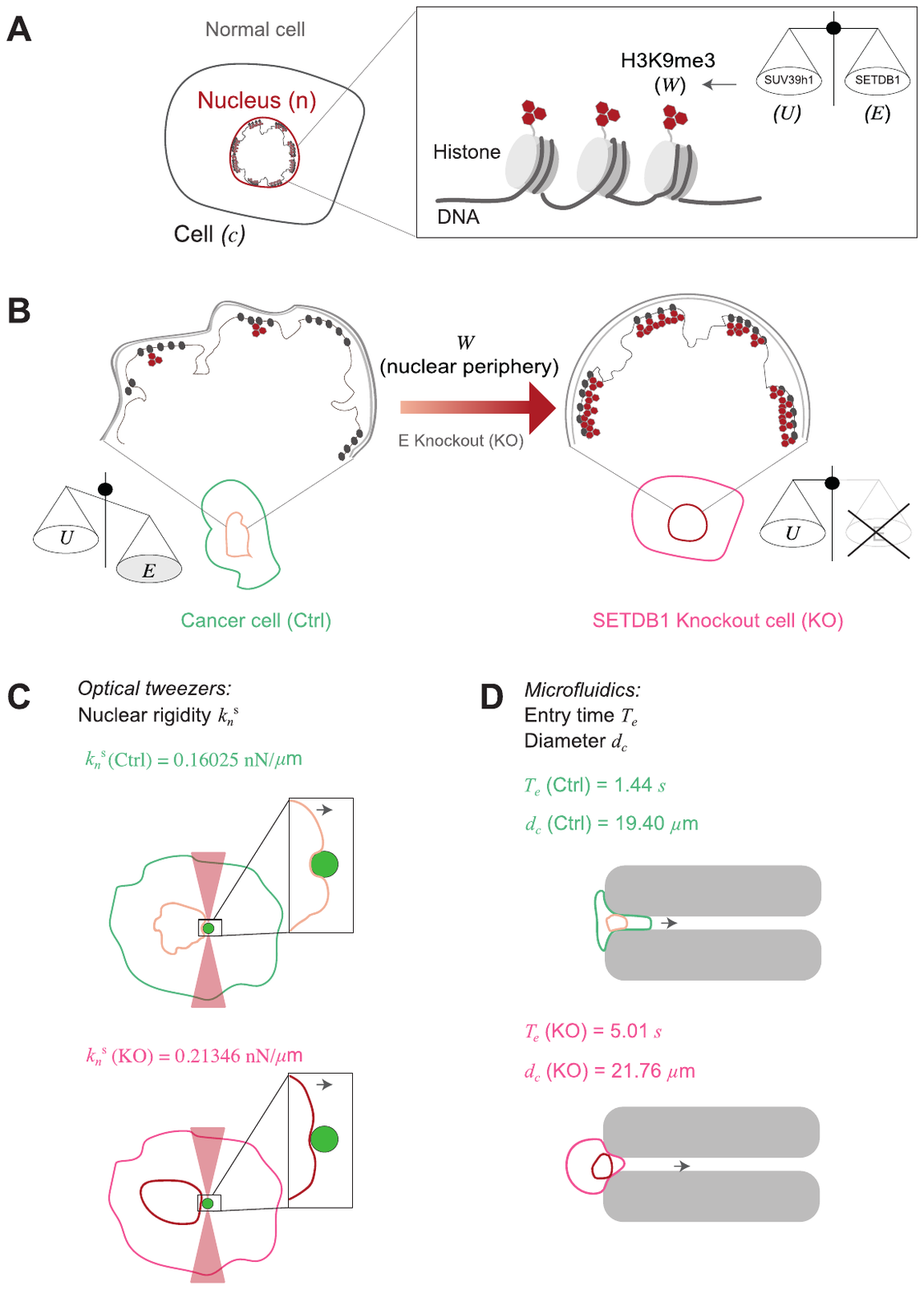}
\caption{\textbf{Schematic overview of the biological system and experimental measurements used to parameterize the mathematical model (adapted from Ref.~\cite{Comunian:2025}).} \\
\textbf{A}. Schematic representation of a cell, $c$ and its nucleus, $n$. Nuclear DNA is packaged around histones into chromatin. H3K9 trimethylation (H3K9me3) is a histone modification associated with heterochromatin formation and is enriched at the nuclear periphery. In the mathematical model, peripheral H3K9me3 is represented by $W$. H3K9me3 distribution is regulated by the interplay between the H3K9 methyltransferases SUV39H1 (denoted $U$ mathematically) and SETDB1 (denoted $E$ mathematically). \\
\textbf{B}. Schematic representation of the experimental system used to modulate peripheral H3K9me3. Control A549 lung cancer cells display relatively low H3K9me3 enrichment at the nuclear periphery due to the dysregulation of SETDB1 ($E$). Deletion of SETDB1 ($E$) alters the balance between SETDB1 and SUV39H1 and promotes SUV39H1-dependent accumulation of H3K9me3 ($W$) at the nuclear periphery. Control and SETDB1 knockout cells therefore provide two experimental conditions characterized by different levels of peripheral H3K9me3. \\
\textbf{C}. Nuclear mechanical properties were measured using optical tweezers. A microsphere positioned adjacent to the nucleus was held in an optical trap while displacement of the cell toward the trapped bead induced nuclear deformation. The resulting force–deformation response was used to determine nuclear rigidity, ($k^{\mathrm{s}}_n$). SETDB1 knockout (KO) cells, characterized by increased peripheral H3K9me3 thanks to SUV39H1 ($U$), displayed higher nuclear rigidity than control cells. Mean experimental values were $k^{\mathrm{s}}_n = 0.16025 \  \mathrm{nN}/\mu\mathrm{m}$ for control cells ($N=61$) and $k^{\mathrm{s}}_n = 0.21346 \  \mathrm{nN}/\mu\mathrm{m}$ for SETDB1 KO cells ($N=69$), pulled from three independent biological replicates. \\
\textbf{D}. Whole-cell deformation under confinement was independently assessed using a microfluidic assay. Cells were driven by an externally imposed flow into a 6-$\mu$m-wide constriction, and the cell entry time, ($T_e$), was measured together with the initial cell diameter, ($d_c$). SETDB1 KO cells required a longer time to enter the constriction than control cells. Mean experimental values were $T_e = 1.44 \ s$ and $d_c = 19.40 \ \mu \mathrm{m}$ for control cells ($N=345$), compared with $T_e = 5.01 \ s$ and $d_c = 21.76 \ \mu \mathrm{m}$ for SETDB1 KO cells ($N=250$), pulled from three independent biological replicates.}
\label{fig:scheme}
\end{figure*}

In non-small cell lung cancer (NSCLC), SETDB1 overexpression has been associated with increased invasive potential and poor clinical outcomes \cite{Cruz-Tapias:2019b, Zakharova:2022}. In the experimental cancer cell model considered here, high SETDB1 expression is unexpectedly associated with reduced H3K9me3 enrichment at the nuclear periphery, lower nuclear rigidity, and greater cellular deformability \cite{Comunian:2025}. Conversely, SETDB1 knockout (SETDB1 KO) shifts the balance towards SUV39H1-dependent H3K9me3 deposition at the nuclear periphery. This redistribution is associated with increased nuclear rigidity and cellular viscosity and with reduced migration through confined spaces. Thus, experimentally altering the balance between SETDB1 and SUV39H1 affects peripheral H3K9me3 and is associated with changes in nuclear mechanics, measured using optical tweezers, and confined migration, measured using microfluidic devices. The molecular interactions underlying this regulatory pathway are schematically illustrated in Fig.~\ref{fig:scheme}.

We selected this extensively characterized cellular system as the biological basis of our mathematical model because it connects a defined molecular process at the nuclear periphery with measurable changes in nuclear mechanics and whole-cell migration. Moreover, the available optical-tweezers and microfluidic measurements provide experimental parameters for the mechanical behaviour of the nucleus and cell, as well as quantitative data against which the mathematical results can be compared.

To quantitatively bridge the gap between macroscopic spatial confinement and molecular regulation at the nuclear periphery, theoretical frameworks must integrate mechanical deformation with molecular dynamics. While various mathematical models have been proposed to describe cell migration \cite{Giverso:2013, Kumar:2018, Scianna:2012, Ballesteros:2019, Leong:2010, Aubry:2014, Allena:2014}, however, capturing the dynamic interplay between molecular processes beneath the nuclear envelope and overarching cellular mechanics remains a significant modelling and computational challenge. To address this question, we develop a fully coupled mathematical framework in which molecular state variables at the nuclear periphery can modulate nuclear mechanical properties. The plasma membrane and the nuclear envelope are represented as two closed, evolving hypersurfaces carrying their own surface energy, whose motion follows geometric evolution laws combining bending resistance, surface tension, and area/perimeter constraints \cite{Elliott:2012, Barreira:2011, Giverso:2013}. Crucially, to incorporate the molecular component, we formulate a system of surface reaction--diffusion partial differential equations posed on the evolving nuclear envelope. This formulation provides a general framework for describing spatio-temporal processes occurring at the nuclear periphery and can be adapted to different molecular regulatory mechanisms. In the present study, we use a descriptive, phenomenological model inspired by the SETDB1/SUV39H1/H3K9me3 pathway. Molecular changes at the nuclear periphery modulate the mechanical properties of the nucleus. The surface reaction–diffusion system is not intended to reproduce the full complexity of nuclear biochemistry, but rather to illustrate how molecular changes at the nuclear periphery can affect nuclear mechanics.
This framework allows us to investigate how changes in nuclear mechanics contribute to confined cell migration, while distinguishing their contribution from changes in whole-cell mechanics.
To solve this highly non-linear and strongly coupled system, we employ the evolving surface finite element method \cite{Neilson:2011, Elliott:2012, Ryder:2019}. By discretizing the moving boundaries as evolving polygonal manifolds, this approach naturally accommodates the complex geometric changes and large deformations experienced by cells navigating narrow microchannels. Ultimately, this computational framework allows us to evaluate the specific influence of nuclear mechanical properties on the efficiency of confined cell migration.

The paper is organised as follows. In Section~\ref{sec:mathematical_model}, we introduce the continuous mathematical framework, detailing the geometric evolution laws for the cell membranes and the surface reaction--diffusion system governing chromatin dynamics. Section~\ref{sec:numerical_method} is devoted to the numerical implementation, where the evolving surface finite element method and the time-discretization scheme are presented. In Section~\ref{sec:results}, we perform numerical simulations to first calibrate the model against experimental data of control and SETDB1 KO cancer cells. Finally, conclusions and future perspectives are drawn in Section~\ref{sec:conclusions}.

\section{Mathematical model}
\label{sec:mathematical_model}
Cell transport is described here by means of a geometric evolution law acting on two distinct, mutually coupled hypersurfaces, which stand for the cell plasma membrane and the nuclear envelope. Building on the framework introduced in \cite{Elliott:2012, Ryder:2019} and later adapted to this setting in \cite{Ballatore:2025}, the two surfaces move according to per-unit-area force balance equations that combine their intrinsic elastic behaviour with the forces of mutual interaction. 
We denote these two evolving, $C^2$-smooth closed surfaces by $\Gamma_c$ and $\Gamma_n$, the subscripts $c$ and $n$ identifying the cell plasma membrane and the nuclear envelope, respectively. A system of geometric surface partial differential equations (GS-PDEs) is prescribed for each membrane, governing its spatial evolution.

Furthermore, epigenetic modifications occurring beneath the nuclear envelope are known to significantly influence nuclear mechanics. Our objective is to incorporate these dynamics by modelling their evolution directly on the nuclear envelope, denoted as $\Gamma_n$. By prescribing the evolution of these epigenetic markers through a system of surface PDEs, we aim to investigate how the levels of these molecules, namely H3K9me3, modulate the overall mechanical behaviour of the nucleus.
 
The scenario examined here is that of a cell being pushed through a microchannel by an externally applied pressure, so as to match the experimental conditions reported in \cite{Jebane:2023}. Following the framework introduced in \cite{Ballatore:2025}, we summarise the mathematical framework, explaining all the terms which comprise the force balance equations. Since we aim to model cell translocation through a microchannel of reduced height, we adopt a two-dimensional computational framework, and thus the evolving surfaces simplify to two evolving one-dimensional curves. 
\subsection{A geometric surface force balance equation for the evolving cell plasma membrane}
Balancing the forces acting on the cell plasma membrane curve $\Gamma_c$ gives:
\begin{equation}
\label{cell_force_balance}
\omega_c\boldsymbol{V}_c=\left(\mathcal{F}_c^{\mathrm{a}}+\mathcal{F}_c^{\mathrm{s}}+\mathcal{F}_c^{\mathrm{b}} +\mathcal{F}_c^{\mathrm{sr}}+\mathcal{F}_c^{\mathrm{ext}}\right)\boldsymbol{\nu}_c \, 
\end{equation}
Here $\boldsymbol{\nu}_c$ is the outward pointing unit normal to the curve $\Gamma_c$, evaluated at the material point $\boldsymbol{x}(t) \in \Gamma_c (t)$, $\omega_c > 0$ plays the role of a friction coefficient, and the cell plasma membrane's material velocity, $\boldsymbol{V}_c =V_c \, \boldsymbol{\nu}_c$, is assumed to point purely along the normal direction. As every force term is assumed to act along the normal direction, the membrane's motion is, by construction, purely normal as well. Such a choice is common practice in geometric evolution frameworks of this kind, and relies on the idea that any tangential component of the velocity only reparametrises the curve, leaving the membrane's actual shape unaffected. The terms on the right-hand side are described as follows.

To regulate the enclosed cell area, we introduce a conservation model driven by a time-dependent pressure difference, $\lambda_c(t)$, between the intracellular and extracellular environments. The corresponding force enforcing area constraint is given by
\begin{equation}
\mathcal{F}_c^{\mathrm{a}} = \lambda_c(t) \, .
\end{equation}
It penalizes deviations from a reference area, which in our case is the initial one, without enforcing the constraint exactly at all times. Note that $\lambda_c(t)$ does not vary in space, and is defined proportionally to the gap between the cell area at time $t$, $A_c(t)$, and a reference cell area, $A^{\mathrm{ref}}_c$. We can then write:
\begin{equation}
\label{lambda_cont}
\lambda_c(t) = - \Lambda_A \left(A_c(t) - A^{\mathrm{ref}}_c\right) \, ,
\end{equation}
where $\Lambda_A$ is a positive penalization parameter that can be interpreted as an effective bulk modulus. Consequently, neither the cell area nor the nuclear area is expected to stay exactly fixed over the course of translocation. However, while Eq.~\eqref{lambda_cont} regulates the enclosed area, the membrane perimeter can still grow excessively. Biologically, cells accommodate membrane expansion by recruiting lipid reservoirs via vesicle fusion. Since these reservoirs are finite, there is a strict upper limit on how much the membrane can expand. To account for this, we extend Eq.~\eqref{lambda_cont} by introducing a tension-like penalty term that activates only when the current cell perimeter, $P_c(t)$, exceeds a maximum reference value, $P_c^{\mathrm{ref}}$:
\begin{multline}
\label{lambda_peri}
\lambda_c(t, \boldsymbol{x})= - \Lambda_A \left(A_c(t) - A_c^{\mathrm{ref}} \right) \\
 - \Lambda_P \left(P_c(t) - P_c^{\mathrm{ref}} \right)_+ H_c \, .
\end{multline}
Note that $\lambda_c(t, \boldsymbol{x})$ is now space-dependent: the perimeter constraint is weighted by the local curvature, $H_c$, meaning the tension penalty is stronger in highly curved regions. Here $H_c$ is the curvature of $\Gamma_c$, given by the tangential divergence of $\boldsymbol{\nu}_c$, i.e., $H_c = \nabla_{\Gamma_c} \cdot \boldsymbol{\nu}_c = \mathrm{tr}\!\left(\nabla_{\Gamma_c}\boldsymbol{\nu}_c\right)$. Throughout, given a scalar field $\eta$ extended smoothly near $\Gamma_c$, $\nabla_{\Gamma_c}$ and $\Delta_{\Gamma_c}$ stand for its tangential gradient and its Laplace–Beltrami operator, computed respectively as:
\begin{equation}
\nabla_{\Gamma_c} \eta := \nabla \eta - (\nabla \eta \cdot \boldsymbol{\nu}_c) \boldsymbol{\nu}_c \, , \quad \Delta_{\Gamma_c} \eta := \nabla_{\Gamma_c} \cdot \left( \nabla_{\Gamma_c} \eta \right) \, .
\end{equation}

The resistance of the cell boundary to stretching is captured by a surface-energy functional, again specialised to a planar curve:
\begin{equation}
\label{surface_energy}
E_{\mathrm{s}}=\int_{\Gamma_c} k_c^{\mathrm{s}} \, \mathrm{d} \Gamma_c \, ,
\end{equation}
where $k_c^{\mathrm{s}}>0$ represents the surface tension coefficient.
Computing the first variation of the energy functional with respect to a normal displacement $\delta u$ yields $\delta (\mathrm{d}\Gamma_c)=-H_c\, \delta u \, \mathrm{d}\Gamma_c$ \cite{Deckelnick:2005}. 
Consequently, the resulting normal force arising from the minimization of surface energy is given by:
\begin{equation}
\mathcal{F}_c^{\mathrm{s}}=-k_c^{\mathrm{s}}H_c \, .
\end{equation}

Being the basic structural constituent of the cell plasma membrane, the lipid bilayer resists bending deformations; this behaviour is described here through the classical Helfrich elastic model \cite{Helfrich:2014} for the bending energy, restricted to a planar curve:
\begin{equation}
\label{bending_energy}
E_{\mathrm{b}}=\int_{\Gamma_c} \frac{1}{2} k_c^{\mathrm{b}} (H_c-H_c^{\mathrm{ref}})^2 \, \mathrm{d} \Gamma_c \, ,
\end{equation}
where $k_c^{\mathrm{b}}>0$ represents the bending rigidity and $H_c^{\mathrm{ref}}$ denotes the spontaneous curvature of the membrane, which, for a circular geometry, is given by the inverse of the reference radius.
Under an infinitesimal normal displacement $\delta u$, the local curvature varies due to both spatial diffusion and intrinsic geometric coupling according to the relation:
\begin{equation}
\delta H_c = \Delta_{\Gamma_c} (\delta u ) + |\nabla_{\Gamma_c} \boldsymbol{\nu}_c|^2 \delta u \, .
\end{equation}
Here, $|\bullet|$ denotes the Frobenius norm. For a planar curve, $|\nabla_{\Gamma_c} \boldsymbol{\nu}_c|^2$ measures the local variation of the unit normal vector and simplifies exactly to the squared curvature, i.e., $|\nabla_{\Gamma_c} \boldsymbol{\nu}_c|^2=H_c^2$. 
Taking the variation of the bending energy with respect to the shape of the membrane, and integrating by parts, we obtain the following force contribution:
\begin{equation}
\mathcal{F}_c^{\mathrm{b}}= k_c^{\mathrm{b}} \left \{  \Delta_{\Gamma_c} H_c + \frac{1}{2} \left[H_c^2-(H_c^{\mathrm{ref}})^2\right]H_c \right \}\, .
\end{equation}
This formulation accounts for the energetic cost associated with changes in membrane curvature, including the forces arising from the local variation of the arc length.

A self-repulsive force $\mathcal{F}_c^{\mathrm{sr}}$, acting at every point $\boldsymbol{x} \in \Gamma_c$, is introduced to avoid unphysical self-intersection of the cell plasma membrane under large deformations. It penalises, in particular, configurations in which membrane points that are far apart along the contour end up very close in space while pointing in nearly opposite directions. Concretely, at every membrane point $\boldsymbol{x}$ we compare its interaction with all other points $\boldsymbol{y}$ falling within a prescribed distance threshold, and take the self-repulsive force at $\boldsymbol{x}$ to be the largest of these pairwise contributions:
\begin{multline}
\mathcal{F}_c^{\mathrm{sr}}(\boldsymbol{x}) = \max_{\boldsymbol{y} \in \Gamma_c \setminus \{\boldsymbol{x}\}} \bigg\{ \frac{1}{2} k_{\mathrm{sr}} \exp\left[ \kappa_d (\delta_d - \|\boldsymbol{x} - \boldsymbol{y}\|) \right] \\ 
\left[ 1 + \tanh\left( \kappa_n \left( \delta_n - \boldsymbol{\nu}_c(\boldsymbol{x}) \cdot \boldsymbol{\nu}_c(\boldsymbol{y}) \right) \right) \right] \bigg\} \, .
\end{multline}
Here $\boldsymbol{\nu}_c(\boldsymbol{x})$ and $\boldsymbol{\nu}_c(\boldsymbol{y})$ are the outward unit normals at $\boldsymbol{x}$ and $\boldsymbol{y}$, and $\|\bullet\|$ gives the Euclidean distance between the two points. The exponential factor grows as the points draw closer together, with $k_{\mathrm{sr}}$ setting the overall strength of the self-repulsion, $\kappa_d$ tuning how sharply it decays with distance, and $\delta_d$ fixing the distance below which repulsion becomes significant. The bracketed term instead grows when the two normals point in nearly opposite directions, so repulsion is strongest for points that are close and oppositely oriented; $\kappa_n$ sets how sharply this depends on relative orientation, while $\delta_n$ marks the orientation beyond which two normals count as nearly opposing.

The external forces acting on the cell membrane, $\mathcal{F}_c^{\mathrm{ext}}$, are finally taken as the sum of four contributions:
\begin{equation}
\mathcal{F}_c^{\mathrm{ext}}=\mathcal{F}_c^{\mathrm{bar}}+\mathcal{F}_c^{\mathrm{pr}}+\mathcal{F}_c^{\mathrm{rep}}+\mathcal{F}_c^{\mathrm{el}} \, .
\end{equation}

First, a barrier force $\mathcal{F}_c^{\mathrm{bar}}$ \cite{Elliott:2012, Giverso:2023} accounts for the interaction of the cell with fixed external constraints, such as the microchannel walls or other obstacles: it keeps the cell from entering forbidden regions by acting repulsively as the membrane draws near the barrier. For a cell moving through a microchannel of fixed width, the wall interaction is described by an exponential repulsion of the form:
\begin{multline}
\mathcal{F}_c^{\mathrm{bar}}(\boldsymbol{x})
= \int_{\mathcal{B}} k_{\mathrm{bar}} \, \exp\!\big(-\beta \, d_{\mathrm{bar}}(\boldsymbol{x}, \boldsymbol{z})\big)\; \\ \boldsymbol{\nu}_{\mathrm{bar}}(\boldsymbol{z}) \cdot \boldsymbol{\nu}_{c}(\boldsymbol{x}) \, \mathrm{d}\ell \, .
\end{multline}
Here $\boldsymbol{z}$ ranges over the barrier curve $\mathcal{B}$; $k_{\mathrm{bar}}$ sets the amplitude of the interaction and $\beta$ governs how quickly it decays with distance; $d_{\mathrm{bar}}(\boldsymbol{x}, \boldsymbol{z})$ is the Euclidean distance from a membrane point $\boldsymbol{x}\in\Gamma_c$ to a barrier point $\boldsymbol{z}$; $\boldsymbol{\nu}_{\mathrm{bar}}(\boldsymbol{z})$ is the outward unit normal at $\boldsymbol{z}$, and $\mathrm{d}\ell$ denotes the infinitesimal arc length.

A second contribution,$\mathcal{F}_c^{\mathrm{pr}}$, represents the active propulsion driving the cell along the channel; within the geometric PDE framework this is implemented as a fixed macroscopic push, projected onto the membrane's local outward normal. Assuming the cell is driven uniformly along the positive $x$-direction, this normal force contribution is defined as:
\begin{equation}
\mathcal{F}_c^{\mathrm{pr}} = k_{\mathrm{pr}} \, (\boldsymbol{e}_x \cdot \boldsymbol{\nu}_c) \, ,
\end{equation}
with $k_{\mathrm{pr}}$ the constant pressure magnitude and $\boldsymbol{e}_x$ the unit vector along $x$-direction.

A further repulsive force, $\mathcal{F}_c^{\mathrm{rep}}$, ensures that the nucleus stays confined inside the cell plasma membrane, preventing it in particular from crossing the cortical layer. Denoting by $\delta(\boldsymbol{x})$ the shortest distance from a membrane point $\boldsymbol{x} \in \Gamma_c$ to the nuclear envelope, $\delta(\boldsymbol{x})= \min_{\boldsymbol{y}\in \Gamma_n} \|\boldsymbol{x} - \boldsymbol{y}\|$, we set: 
\begin{equation}
\mathcal{F}_c^{\mathrm{rep}}(\boldsymbol{x}) = k_{\mathrm{rep}}\, \phi(\delta(\boldsymbol{x})) \, ,
\end{equation}
where the stiffness parameter $k_{\mathrm{rep}} > 0$ sets how strongly the penalization acts and $\phi$ is a decreasing, positive function of $\delta$, so that the push grows stronger as the gap narrows, defined for example as:
\begin{equation}
  \phi(\delta) = e^{-\alpha\, \delta}, \quad \text{with } \alpha > 0 \, .
\end{equation}

Finally, an elastic restoring force is used to represent the mechanical coupling, mediated by the cytoskeleton, between the cell membrane and the nucleus. This force prevents the cell plasma membrane from drifting too far from the nucleus. Its orientation is defined by the relative displacement between the cell centroid, $\boldsymbol{x}_c$, and the nuclear centroid, $\boldsymbol{x}_n$. Since the geometric framework accounts only for the normal components of the forces, this interaction is projected onto the local unit normal vector of the cell plasma membrane. The resulting force is expressed as:
\begin{equation}
\mathcal{F}_{c}^{\mathrm{el}} = - {k}_{\mathrm{el}} \left(\boldsymbol{x}_c - \boldsymbol{x}_n\right) \cdot \boldsymbol{\nu}_c \, ,
\end{equation}  
where ${k}_{\mathrm{el}} > 0$ denotes the effective elastic dragging coefficient.

Summing every force term introduced above produces the fourth-order GS-PDE that determines how the cell plasma membrane evolves:
\begin{multline}
\label{eq:cellular_balance}
\omega_c \boldsymbol{V}_c = \bigg\{ -k_c^{\mathrm{s}}H_c + k_c^b\Delta_{\Gamma_c}H_c \\ +\frac{1}{2}k_c^b \left[ H_c^2-(H_c^{\mathrm{ref}})^2\right]H_c 
 + \lambda_c(t, \boldsymbol{x}) \\ + \mathcal{F}_c^{\mathrm{sr}} + \mathcal{F}_c^{\mathrm{bar}}+\mathcal{F}_c^{\mathrm{pr}}+\mathcal{F}_c^{\mathrm{rep}}+\mathcal{F}_c^{\mathrm{el}}  \bigg\} \boldsymbol{\nu}_c \, .
\end{multline}

\subsection{A geometric surface force balance equation for the evolving nuclear envelope}
We now complete the formulation with the force balance equation governing the nuclear envelope. Similar to the cell membrane, the nuclear envelope is subject to intrinsic forces arising from area conservation, viscous resistance, perimeter tension, and bending stiffness, as well as external interactions. Although the governing geometric surface PDE has the same mathematical structure as for the cell plasma membrane, its material parameters are chosen so as to reflect the markedly greater stiffness and mechanical resistance of the nucleus reported experimentally \cite{Lammerding:2011}. Moreover, the nuclear mechanical parameters $\omega_n$, $k^{\mathrm{s}}_n$, and $k^{\mathrm{b}}_n$ depend dynamically on the molecular state variables introduced in Section~\ref{sec:RD-PDE}, as detailed in Section~\ref{sec:functions}.

In the normal direction, the total force balance on the nuclear envelope $\Gamma_n$ reads:
\begin{equation}
\label{eq:nuclear_balance}
\omega_n \boldsymbol{V}_n = \left(\mathcal{F}_n^{\mathrm{a}} + \mathcal{F}_n^{\mathrm{s}} + \mathcal{F}_n^{\mathrm{b}} + \mathcal{F}_n^{\mathrm{sr}} + \mathcal{F}_n^{\mathrm{ext}}\right) \boldsymbol{\nu}_n \, ,
\end{equation}
with $\boldsymbol{\nu}_n$ the outward unit normal on $\Gamma_n$ and $\boldsymbol{V}_n$ its material velocity. The external force contribution $\mathcal{F}_n^{\mathrm{ext}}$ decomposes into repulsive and elastic components:
\begin{equation}
\label{eq:nuclear_external}
\mathcal{F}_n^{\mathrm{ext}} =  \mathcal{F}_n^{\mathrm{rep}} + \mathcal{F}_n^{\mathrm{el}} \, .
\end{equation}
Here, the repulsive force $\mathcal{F}_n^{\mathrm{rep}}$ and the elastic force $\mathcal{F}_n^{\mathrm{el}}$ represent the coupling interactions between the nucleus and the cell membrane (via the cytoskeleton and cytoplasmic fluid). The nucleus is not subject to a direct confinement force from the channel; instead, its confinement is mediated indirectly through its mechanical coupling with the cell membrane, in particular through the repulsive interaction $\mathcal{F}_n^{\mathrm{rep}}$. Consequently, these terms share the same functional form as their counterparts acting on the cell membrane but appear with an opposite sign, ensuring the balance of internal action--reaction forces.

\subsection{Derivation of the reaction-diffusion equations on an evolving surface}
\label{sec:RD-PDE}
Beyond the core biophysical mechanisms described above, the mathematical model is further extended to incorporate the chromatin--mechanical regulatory process outlined in the Introduction. In the present study, we derive the general framework to describe the SETDB1/SUV39H1/H3K9me3 pathway and its role in regulating nuclear mechanics. The molecular mechanism considered here and its connection to nuclear stiffness are schematically summarised in Fig.~\ref{fig:scheme}.

These effects are incorporated into the modelling framework by introducing additional molecular state variables describing chromatin organisation at the nuclear envelope \cite{Madzvamuse:2016, Cusseddu:2018}. In this modelling framework, these variables are defined directly on the evolving nuclear surface $\Gamma_n(t)$ and governed by surface reaction--diffusion partial differential equations (S-RDEs) posed on the nuclear envelope. This surface-based description captures the dominant regulatory role of dynamically deforming lamina-associated heterochromatin in nuclear mechanics while preserving a parsimonious mathematical formulation. 

Following the derivation in \cite{Barreira:2011, Elliott:2012, Madzvamuse:2016}, let $S$ denote a generic dimensionless molecular state variable defined on the evolving nuclear surface $\Gamma_n(t)$. The corresponding conservation law states that the rate of change of the total amount of the molecular state variable $S$ within an evolving material region is balanced by the outward flux $\boldsymbol{q}$ across its boundary together with a net reactive production term $F$. Let $\mathcal{R}(t)$ be an arbitrary material subregion of $\Gamma_n(t)$ whose points move with the material velocity. According to the conservation of mass principle:
\begin{equation}
\dfrac{\mathrm{d}}{\mathrm{d}t} \int_{\mathcal{R}(t)} S = - \int_{\partial\mathcal{R}(t)} \boldsymbol{q}\cdot \boldsymbol{\nu}_n +  \int_{\mathcal{R}(t)} F \, .
\end{equation}
Applying the Reynolds transport theorem on evolving surfaces \cite{Dziuk:2007b} to the left-hand side, together with integration by parts on the flux term, yields
\begin{equation}
\int_{\mathcal{R}(t)} \partial^{\bullet}S + S\nabla_{\Gamma_n} \cdot \boldsymbol{V}_n = - \int_{\mathcal{R}(t)} \nabla_{\Gamma_n}\cdot \boldsymbol{q} +  \int_{\mathcal{R}(t)} F \, .
\end{equation}
Here, $\partial^{\bullet}(\cdot) = \partial_t(\cdot) + \boldsymbol{V}_n \cdot \nabla_{\Gamma_n(t)}(\cdot)$ represents the material derivative and $\boldsymbol{V}_n$ is the corresponding material velocity as defined earlier by the geometric surface PDE on the nuclear envelope. Since $\mathcal{R}(t)$ is arbitrary, localisation gives
\begin{equation}
\partial^{\bullet}S + S\nabla_{\Gamma_n} \cdot \boldsymbol{V}_n = \mathsf{D}_S\Delta_{\Gamma_n} S +  F \, ,
\end{equation}
where we assume Fickian diffusion according to $\boldsymbol{q}=-\mathsf{D}_S \nabla_{\Gamma_n}S$. 

To model the molecular signalling pathway described at the beginning of the section, let $E$, $U$, and $W$ denote the dimensionless molecular state variables associated with SETDB1, SUV39H1, and H3K9me3, respectively. They provide a phenomenological representation of the regulatory interactions relevant to the model and should not be interpreted as direct measurements of molecular concentrations. The corresponding system of S-RDEs on $\Gamma_n(t)$ is given by
\begin{align}
\label{eq:PDEs}
\begin{cases}
\partial^{\bullet}E + E\nabla_{\Gamma_n} \cdot \boldsymbol{V}_n = \mathsf{D}_E\Delta_{\Gamma_n} E + f(E,U,W) \\
\partial^{\bullet}U + U\nabla_{\Gamma_n} \cdot \boldsymbol{V}_n = \mathsf{D}_U\Delta_{\Gamma_n} U + g(E,U,W) \\
\partial^{\bullet}W + W\nabla_{\Gamma_n} \cdot \boldsymbol{V}_n = \mathsf{D}_W\Delta_{\Gamma_n} W + h(E,U,W) \, .
\end{cases}
\end{align}
Following a phenomenological approach based on the law of mass action, we propose the following surface reaction terms to model the interplay described in Fig.~\ref{fig:scheme}:
\begin{align}
f(E,U,W) &= k_1E\left(1-E\right) - aE \, , \\
g(E,U,W) &= k_2U\left(1-U\right) - k_3E - bU \, , \\
h(E,U,W) &= k_4 U - cW \, .
\end{align}
The molecular state variables $E$ and $U$ exhibit logistic growth together with linear degradation terms. Since these state variables are dimensionless and normalised, logistic kinetics provide a simple phenomenological mechanism to account for saturation effects and to maintain the variables within a bounded range. Additionally, the inhibitory effect of SETDB1 on SUV39H1 is represented through the linear term $-k_3E$. This formulation assumes that SETDB1 modulates the effective activity of SUV39H1 rather than directly altering the underlying SUV39H1 state. This choice provides a minimal phenomenological description of the experimentally observed antagonistic regulation. Finally, the production of H3K9me3 is modelled through the linear term $k_4U$, reflecting the assumption that SUV39H1 availability controls H3K9me3 accumulation at the nuclear envelope through a first-order approximation.

The regulatory network is formulated as a feed-forward pathway, consistent with the experimentally described SETDB1--SUV39H1--H3K9me3 regulatory cascade. Potential feedback mechanisms involving H3K9me3 are not considered here, as their role in this specific regulatory context has not been characterised and would require additional experimental investigation. Overall, the proposed reaction system is intended as a minimal phenomenological representation of the main regulatory mechanisms affecting nuclear envelope properties, rather than a detailed reconstruction of the complete underlying molecular network.

To characterize the long-term behaviour of the system, we first investigate the spatially homogeneous steady states. The system admits the following trivial and semi-trivial equilibria:
\begin{align}
\left( E^*_1, U^*_1, W^*_1 \right) &= \left( 0 \, , \, 0 \, , \, 0 \right) \, , \\
\left( E^*_2, U^*_2, W^*_2 \right) &= \left( 0 \, , \frac{(k_2-b)}{k_2} ,\frac{k_4(k_2-b)}{ck_2} \right) .
\end{align}
The second equilibrium point is physically feasible (i.e., dimensionless molecular state variables are non-negative) if and only if $k_2 \ge b$. 
Furthermore, provided that the discriminant 
\begin{equation}
\label{delta}
\Delta = (k_2-b)^2 - 4\dfrac{k_2k_3}{k_1}(k_1-a)
\end{equation}
is non-negative, the system exhibits two additional equilibria, $(E^*_3, U^*_3, W^*_3)$ and $(E^*_4, U^*_4, W^*_4)$, where:
\begin{align}
\left( E^*_3, U^*_3, W^*_3 \right) &= \\ \nonumber & \hspace{-10mm} \left( \frac{(k_1-a)}{k_1}, \frac{\left[(k_2-b)-\sqrt{\Delta}\right]}{2k_2}, \frac{k_4U^*_3}{c} \right ) \, , \\
\left( E^*_4, U^*_4, W^*_4 \right) &= \\ \nonumber & \hspace{-10mm} \left( \dfrac{(k_1-a)}{k_1}, \dfrac{\left[(k_2-b)+\sqrt{\Delta}\right]}{2k_2}, \dfrac{k_4U^*_4}{c} \right) \, .
\end{align}
The third and fourth equilibria are physically feasible if and only if the condition $k_1 \ge a$ is satisfied. Hereafter, we assume that all these feasibility conditions are satisfied, namely, $k_1 \ge a$, $k_2 \ge b$, and $\Delta \ge 0$. The stability of these states is governed by the Jacobian matrix:
\begin{equation}
J = \begin{pmatrix}
k_1\left( 1-{2E}\right)-a & 0 & 0 \\
-k_3 & k_2\left( 1-{2U}\right)-b & 0 \\
0 & k_4 & -c \\
\end{pmatrix} 
\end{equation}
Since $J$ is lower triangular, its eigenvalues correspond to the diagonal entries. It is straightforward to verify that for the first three equilibria, at least one eigenvalue is positive, rendering them unstable. In contrast, the fourth equilibrium is locally stable as all diagonal entries are negative. Notably, the global asymptotic stability of $(E^*_4, U^*_4, W^*_4)$ follows from the cascade structure of the system: the equation for $E$ is a decoupled logistic-like ODE that converges globally to $E^*_4$ for any positive initial condition. This convergence, in turn, sequentially drives $U$ and $W$ to their respective uniform steady states.

When considering the spatial domain, the lower triangular structure of the Jacobian ensures that diffusion acts exclusively as a stabilizing agent. Consequently, Turing instabilities are strictly precluded, and the system naturally relaxes toward a spatially homogeneous steady state. This is numerically confirmed by considering both spatially constant initial conditions, e.g., 
\begin{equation}
\label{IC_const}
(E, U, W)|_{t=0} = (0.2 \, , \, 0.2 \, , \, 0) \, ,  
\end{equation}
and non-constant perturbations of the form 
\begin{multline}
\label{IC_cos}
(E, U, W)|_{t=0} = \\ \ (0.2 + 0.2\cos(10 \phi) \, , \,  0.2 + 0.2\cos(10 \phi) \, ,0) \, ,
\end{multline}
where $\phi = \arctan(y/x)$. In both cases, the system converges to $(E^*_4, U^*_4, W^*_4)$ as $t \to \infty$. Conversely, in the SETDB1 KO scenario, where the molecular state variable $E$ is set to zero on the nuclear envelope ($E(0,\boldsymbol{x}) \equiv 0$), the system converges to $(E^*_2, U^*_2, W^*_2)$, resulting in a higher asymptotic level of $W$.

Finally, chromatin dynamics are coupled to the mechanical evolution by allowing parameters such as friction-like coefficient ($\omega_n$), surface tension ($k^{\mathrm{s}}_n$) and bending rigidity ($k^b_n$) to depend explicitly on the local H3K9me3 value, i.e., $\omega_n=f_1(W)$, $k^{\mathrm{s}}_n=f_2(W)$ and $k^b_n=f_3(W)$. The specific functional forms of $f_1$, $f_2$, and $f_3$, derived from comparison with experimental data, are detailed in the following section to describe the coupling between the molecular state variables and the corresponding mechanical responses.

\subsection{Coupling nuclear mechanics to dimensionless molecular state variables}
\label{sec:functions}
To couple the molecular state variable $W$ beneath the nuclear envelope $\Gamma_n$ with the mechanical response of the system, we define the friction-like coefficient $\omega_n$, the surface tension $k^{\mathrm{s}}_n$, and the bending stiffness $k^b_n$, as functions of the local value of $W$, denoted by $f_1(W)$, $f_2(W)$ and $f_3(W)$, respectively. 

To define the dynamic nuclear friction coefficient $\omega_n = f_1(W)$, we rely on the phenomenological observation that control (Ctrl) cells migrate faster than SETDB1 KO cells, implying a lower effective friction for the control phenotype. As observed in Section~\ref{sec:RD-PDE}, the asymptotic value of the molecular state variable $W$ is  lower in Ctrl cells ($W_4^*$) compared to SETDB1 KO cells ($W_2^*$). To reflect this positive correlation between the nuclear friction and the availability of $W$, we define $f_1(W)$ to scale proportionally with the spatial average of $W$ over the nuclear envelope. Thus, we propose the following thresholded linear relationship for the friction-like coefficient:
\begin{equation}
\omega_n = f_1(W) = \max\left(\omega_c \langle W \rangle_{\Gamma_n}, \frac{1}{2}\omega_c\right),
\end{equation}
where
\[
\langle W \rangle_{\Gamma_n} = \frac{1}{|\Gamma_n|} \int_{\Gamma_n} W \, \mathrm{d}\ell
\]
denotes the instantaneous spatial average of the H3K9me3 value on the nuclear envelope, and $\omega_c$ represents the reference friction scale associated with the cell membrane. This formulation introduces a lower bound for the nuclear friction coefficient, preventing it from decreasing below $\frac{1}{2}\omega_c$. Specifically, the friction coefficient follows the linear dependence $\omega_n=\omega_c\langle W\rangle_{\Gamma_n}$ when the averaged H3K9me3 value is sufficiently large, whereas it is maintained at the minimum value $\frac{1}{2}\omega_c$ otherwise. 

Furthermore, experimental measurements obtained using optical tweezers provide the nuclear mechanical properties of Ctrl and SETDB1 KO cells, from which we determine the reference nuclear surface tension values $\hat{k}^{\mathrm{s}}_n$ for Ctrl cells and $\tilde{k}^{\mathrm{s}}_n$ for SETDB1 KO cells. For the surface tension function $f_2(W)$, we assume a monotonically increasing dependence on the value of the molecular state variable $W$, interpolating between the two reference mechanical states associated with Ctrl and SETDB1 KO cells. Specifically, the function is required to satisfy
\begin{equation}
    f_2(W_4^*) = \hat{k}^{\mathrm{s}}_n, 
    \qquad
    f_2(W_2^*) = \tilde{k}^{\mathrm{s}}_n .
\end{equation}
A thresholded linear interpolation is then introduced:
\begin{equation}
\label{f2}
    f_2(W) =
    \dfrac{\tilde{k}^{\mathrm{s}}_n-\hat{k}^{\mathrm{s}}_n}
    {W^*_2-W^*_4}
    (W-W_4^*)_+
    +\hat{k}^{\mathrm{s}}_n ,
\end{equation}
where $(\cdot)_+$ denotes the positive part. This formulation ensures that, for H3K9me3 levels below the control reference value $W_4^*$, the nuclear surface tension remains equal to the control value $\hat{k}^{\mathrm{s}}_n$. For increasing values of $W$, the surface tension increases linearly until reaching the SETDB1 KO reference value $\tilde{k}^{\mathrm{s}}_n$ at $W=W_2^*$.

Finally, a similar interpolation strategy is adopted for the bending stiffness of the nuclear envelope, which is governed by the function $f_3(W)$. Starting from the same baseline cellular bending stiffness $k^b_c$ for both Ctrl and SETDB1 KO cells, we define the nuclear bending stiffness as $\hat{k}^b_n = 10k^b_c$ for Ctrl cells and $\tilde{k}^b_n = 100k^b_c$ for SETDB1 KO cells, thereby accounting for the greater structural rigidity of the nucleus relative to the plasma membrane and its further increase in the SETDB1 KO condition. We assume $f_3(W)$ to be a monotonically increasing relationship that satisfies the following conditions:
\begin{equation}
    f_3(W_4^*) = \hat{k}^b_n \, , \quad f_3(W_2^*) = \tilde{k}^b_n \, .
\end{equation}
The resulting linear interpolation formula is expressed as:
\begin{equation}
\label{f3}
    f_3(W) = \dfrac{\tilde{k}^b_n-\hat{k}^b_n}{W^*_2-W^*_4}\left(W-W_4^* \right)_+ +\hat{k}^b_n  \, .
\end{equation}

\section{Numerical implementation}
\label{sec:numerical_method}
The objective of this section is to describe the numerical implementation of the mathematical framework presented in Section~\ref{sec:mathematical_model}, specifically the coupled system of equations \eqref{eq:cellular_balance}, \eqref{eq:nuclear_balance}, and \eqref{eq:PDEs}. We first derive the weak formulation of the governing equations, which provides the foundation for the surface finite element treatment. Subsequently, we detail the spatial discretization and the temporal integration scheme employed to solve the evolution of both the mechanical configuration and the molecular state variables. Finally, the computational implementation and the solution of the resulting discrete algebraic systems are carried out using the open-source finite element framework FEniCSx.
\subsection{Weak formulation}
In this section, we derive the variational formulation for the geometric evolution of the membranes and the surface reaction--diffusion system. For the sake of brevity, we present a generalized formulation applicable to both the plasma membrane and the nuclear envelope, omitting specific superscripts.

A fundamental identity underlying the numerical method is the following relation between the Laplace--Beltrami operator of the identity map and the mean curvature vector:
\begin{equation}
\Delta_{\Gamma} \boldsymbol{x} = H \boldsymbol{\nu} \, .
\end{equation}
The derivation follows a Lagrangian approach, where $\boldsymbol{x}(t)$ denotes the position of a material point on the surface. We seek the position $\boldsymbol{x} \in \mathrm{H}^{1}(\Gamma(t))^2$ and the curvature $H \in \mathrm{H}^{1}(\Gamma(t))$ such that:
\begin{equation}
\label{eq:weak_mech}
\begin{cases}
\displaystyle \int_{\Gamma(t)} (
   \omega \, \partial_t \boldsymbol{x} \cdot \boldsymbol{\nu} \, \chi 
   + k_b \, \nabla_{\Gamma} H \cdot \nabla_{\Gamma} \chi 
    \\ \hspace{22mm} - \frac{1}{2} k_b(H^2-H^2_{\mathrm{ref}})H\chi
   + k_s H \chi) \, \mathrm{d}\ell \\
   \hspace{5mm}= \displaystyle \int_{\Gamma(t)} \left( \lambda(t, \boldsymbol{x}) + \mathcal{F}_{\mathrm{sr}} 
    + \mathcal{F}_{\mathrm{ext}} \right) \chi \, \mathrm{d}\ell \, , \\
\displaystyle \int_{\Gamma(t)}  \left( H \, \boldsymbol{\nu} \cdot \Phi 
   + \nabla_{\Gamma} \boldsymbol{x} : \nabla_{\Gamma} \Phi  \right) \, \mathrm{d}\ell 
   = 0 \, ,
\end{cases}
\end{equation}
for all test functions $\chi \in \mathrm{H}^{1}(\Gamma(t))$ and $\Phi \in \mathrm{H}^{1}(\Gamma(t))^2$. In \eqref{eq:weak_mech}, $\boldsymbol{\nu}$ is the outward unit normal, ``$:$'' denotes the Frobenius inner product between two tensors and we have utilized the Divergence Theorem on surfaces. The absence of boundary terms is guaranteed by the closed nature of the membranes.

Regarding the molecular state variables on the evolving nuclear envelope $\Gamma_n(t)$, the weak formulation is obtained by applying the Transport Theorem, following a reasoning similar to that developed in \cite{Barreira:2011}. The system of S-RDEs then reads:
\begin{align}
& \dfrac{\mathrm{d}}{\mathrm{d}t} \int_{\Gamma_n(t)} E \, \varphi \, \mathrm{d}\ell + \int_{\Gamma_n(t)} D_E \nabla_{\Gamma_n} E \cdot \nabla_{\Gamma_n} \varphi \, \mathrm{d}\ell \nonumber \\
& = \int_{\Gamma_n(t)} E \partial^{\bullet} \varphi + \int_{\Gamma_n(t)} f(E,U,W) \, \varphi \, \mathrm{d}\ell \, , \\[6pt]
&\dfrac{\mathrm{d}}{\mathrm{d}t} \int_{\Gamma_n(t)} U \, \varphi \, \mathrm{d}\ell + \int_{\Gamma_n(t)} D_U \nabla_{\Gamma_n} U \cdot \nabla_{\Gamma_n} \varphi \, \mathrm{d}\ell \nonumber \\
& = \int_{\Gamma_n(t)} U \partial^{\bullet} \varphi + \int_{\Gamma_n(t)} g(E,U,W) \, \varphi \, \mathrm{d}\ell \, , \\[6pt]
&\dfrac{\mathrm{d}}{\mathrm{d}t} \int_{\Gamma_n(t)} W \, \varphi \, \mathrm{d}\ell + \int_{\Gamma_n(t)} D_W \nabla_{\Gamma_n} W \cdot \nabla_{\Gamma_n} \varphi \, \mathrm{d}\ell \nonumber \\
& = \int_{\Gamma_n(t)} W \partial^{\bullet} \varphi + \int_{\Gamma_n(t)} h(E,U,W) \, \varphi \, \mathrm{d}\ell \, ,
\end{align}
for all $\varphi \in \mathrm{H}^1(\Gamma_n(t))$. This formulation implicitly accounts for the surface dilution effect appearing in the strong form, which is here absorbed into the time derivative of the integral.

\subsection{Spatial and temporal discretization}
\label{sec:discretization}
A spatial and temporal discretization of all the equations involved is performed based on evolving surface finite elements \cite{Barreira:2011, Barrett:2008, Dziuk:2013}. The time interval $[0,T]$ is partitioned into $M$ subintervals, $0 < t^1 < \dots < t^M = T$, of constant length $\tau$. 

The method relies on a time-discrete sequence of evolving surfaces ${\Gamma_h^m}$, where $\Gamma_h^m:=\Gamma_h(t^m)$ denotes the discrete surface at time $t^m$. Given a triangulation $\mathcal{T}_h^m$ approximating $\Gamma_h^m$, we define the fully discrete finite element space of globally continuous piecewise quadratic polynomials by
\begin{equation}
\mathbb{V}_h^m := \left\{ \chi \in C(\Gamma_h^m) : \ \chi|_s \ \text{is quadratic}, \ \forall s \in \mathcal{T}_h^m \right\}.
\end{equation}

For the mechanical evolution, given the previous states $\boldsymbol{x}_h^m \in (\mathbb{V}_h^m)^2$, $H_h^m \in \mathbb{V}_h^m$, and the molecular state variable $W_h^m \in \mathbb{V}_h^m$, we seek the updated position $\boldsymbol{x}_h^{m+1} \in (\mathbb{V}_h^{m+1})^2$ and curvature $H_h^{m+1} \in \mathbb{V}_h^{m+1}$ such that for $m = 0, \dots, M-1$:
\begin{equation}
\label{FEM1}
\left\{
\begin{aligned}
& \int_{\Gamma_h^m} \Bigg( \dfrac{\omega(W_h^m)}{\tau} (\boldsymbol{x}_h^{m+1} - \boldsymbol{x}_h^m) \cdot \boldsymbol{\nu}_h^m \, \chi_h \\
& \quad + k_b(W_h^m) \, \nabla_{\Gamma_h^m} H_h^{m+1} \cdot \nabla_{\Gamma_h^m} \chi_h \\
& \quad - \dfrac{1}{2} k_b(W_h^m) \, \left[ \left(H_h^{m+1}\right)^2 - H^2_{\mathrm{ref}}\right]\, H_h^{m+1} \, \chi_h \\
& \quad + k_s(W_h^m) \, H_h^{m+1} \, \chi_h \Bigg) \, \mathrm{d}\ell \\
& \quad = \int_{\Gamma_h^m} \big( \lambda_h^{m+1} + \mathcal{F}_{\mathrm{sr}}^m + \mathcal{F}_{\mathrm{ext}}^m \big) \, \chi_h \, \mathrm{d}\ell \, , \\[10pt]
& \int_{\Gamma_h^m} \bigg( H_h^{m+1} \, \boldsymbol{\nu}_h^m \cdot \Phi_h + \\
& \hspace{1.5cm} \nabla_{\Gamma_h^m} \boldsymbol{x}_h^{m+1} : \nabla_{\Gamma_h^m} \Phi_h \bigg) \, \mathrm{d}\ell = 0 \, .
\end{aligned}
\right.
\end{equation}
Here, the resulting semi-implicit backward Euler scheme is solved here through a Picard-type iterative procedure \cite{Madzvamuse:2006, Madzvamuse:2014}.

It is worth highlighting that, for the evolving nuclear envelope $\Gamma_n$, the biomechanical coupling is treated explicitly. Consequently, the mechanical parameters of the nucleus depend on the value of the molecular state variable $W$ evaluated at the previous time step, yielding $\omega_n=f_1(W_h^m)$, $k_n^{\mathrm{s}}=f_2(W_h^m)$, and $k_n^b=f_3(W_h^m)$, while the corresponding parameters of the cell plasma membrane are kept constant, namely $\omega_c$, $k_c^{\mathrm{s}}$, and $k_c^b$.

Furthermore, the discrete geometric conservation of the cell area is enforced by updating $\lambda_h^{m+1}$ via:
\begin{multline}
\label{lambda_m}
\lambda_h^{m+1} = - \Lambda_A \left(A^{m+1} - A_{\mathrm{ref}} \right) \\
 - \Lambda_P \left(P^{m+1} - P_{\mathrm{ref}} \right)_+ H_h^{m+1} \, ,
\end{multline}
where $A_{\mathrm{ref}}$ denotes the reference initial area, $A^{m+1}$ is the discrete area at time step $m+1$, while $P_{\mathrm{ref}}$ denotes the reference initial perimeter and $P^{m+1}$ is the discrete perimeter at time step $m+1$.

Finally, we present the fully discrete formulation for the S-RDEs governing the molecular state variable $S \in \{E, U, W\}$ on the evolving nuclear envelope $\Gamma_{n,h}$. Consistent with the transport theorem underlying the continuous weak form, the mass accumulation term gives rise to the difference between the integrals evaluated on the current and previous discrete surfaces. Let $\left\{ \varphi_j(\cdot, t) \right\}_{j=1}^N$ denote the moving nodal basis functions satisfying the discrete transport property $\partial^{\bullet}_h\varphi_j = 0$ \cite{Barreira:2011}, we seek $S_h^{m+1} \in \mathbb{V}^{m+1}$ such that:
\begin{equation}
\label{FEM2}
\begin{aligned}
& \frac{1}{\tau} \int_{\Gamma_{n,h}^{m+1}} S_h^{m+1} \varphi^{m+1}_j \, \mathrm{d}\ell 
\\
&\qquad  +\int_{\Gamma_{n,h}^{m+1}} \mathsf{D}_S \nabla_{\Gamma_{n,h}^{m+1}} S_h^{m+1} \cdot \nabla_{\Gamma_{n,h}^{m+1}} \varphi^{m+1}_j \, \mathrm{d}\ell \\
& \qquad = \dfrac{1}{\tau} \int_{\Gamma_{n,h}^m} S_h^m \varphi^m_j \, \mathrm{d}\ell + \int_{\Gamma_{n,h}^{m+1}} F^{m+1} \varphi^{m+1}_j \, \mathrm{d}\ell \, ,
\end{aligned}
\end{equation}
for $j=1,\ldots,N$, where $\mathsf{D}_S$ denotes the diffusion coefficient associated with molecular state variable $S$. The nonlinear reaction functions $F^{m+1}=F(E_h^{m+1},U_h^{m+1},W_h^{m+1})$, with $F\in\{f,g,h\}$, are evaluated implicitly, yielding a stable coupling between the molecular dynamics and the geometric evolution.

The numerical algorithm for solving \eqref{FEM1} and \eqref{FEM2} was implemented using FEniCSx \cite{Baratta:2023}. To ensure the stability of the GS-PDE, a suitable distribution of the mesh nodes is maintained by introducing a tangential component into the velocity \cite{Elliott:2012, Barrett:2020, Barrett:2008b}. Since tangential motion only redistributes the nodes along the curve without altering its geometry, the force balance remains unchanged. The corresponding tangential displacement is computed using the De Boor algorithm adapted to closed curves \cite{deBoor:2006, Hernández:2024}.

\subsection{Parameter selection}
\label{sec:parameters}
The accurate identification and estimation of numerical parameters are crucial for the calibration of the model. As several of these biophysical quantities are not directly measurable, their values are typically approximated through indirect estimation or inferred from \textit{in vitro data}. Here, we adopt parameter values already established in prior experimental work and simulation studies, drawing especially on the results in \cite{Jebane:2023, Comunian:2025}. A comprehensive summary of the selected parameter values, along with their literature references, is provided in Table \ref{tab:parameters}. 

\begin{figure}[t]
    \centering
    \includegraphics[width=\linewidth]{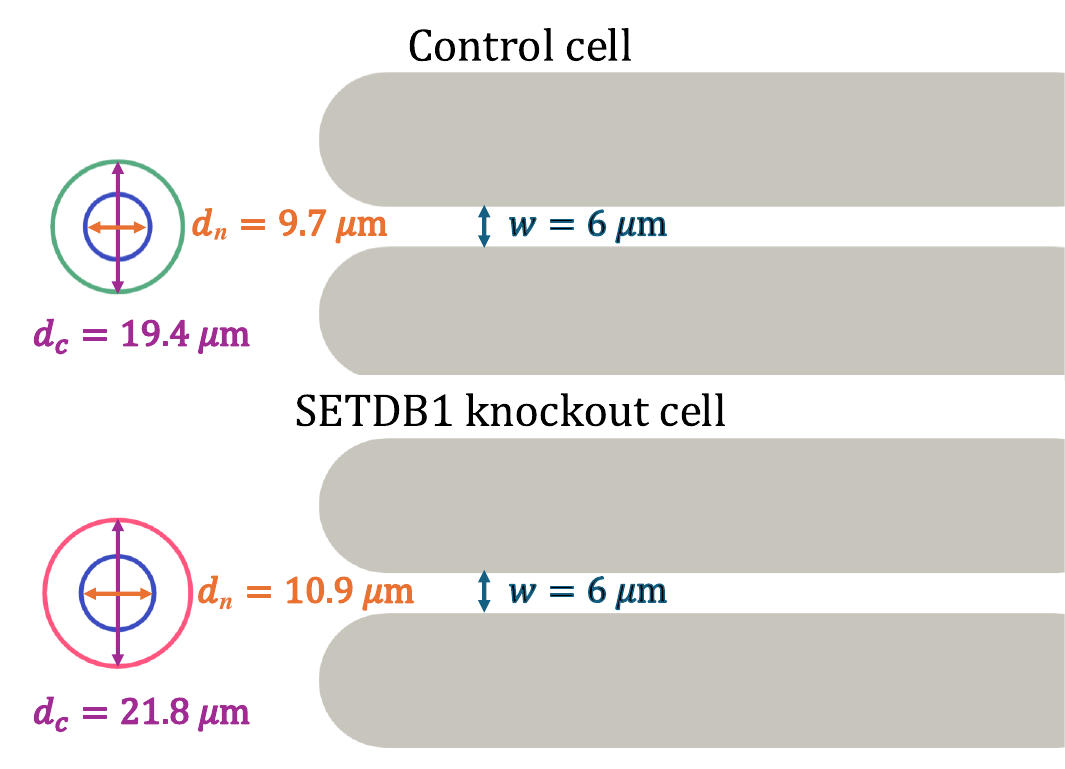}
    \caption{Initial computational configuration adopted in the simulations. The geometry is shown to scale, with the relative dimensions of the cell, nucleus, and microchannel preserved. The blue circle denotes the initial position of the nucleus, while the green and red circles denote the initial positions of the Ctrl and SETDB1 KO cells, respectively. The corresponding diameters are indicated for each configuration.}
    \label{fig:setup}
\end{figure}

Regarding the baseline dimensions of the cell, we rely on experimental measurements obtained by the authors through microfluidic experiments. Accordingly, for the Ctrl cells, we set the initial cell diameter to $d_c = 19.4 \ \mu\text{m}$ and the nuclear diameter estimated to half of this value, yielding $d_n = 9.7 \ \mu\text{m}$. Similarly, for the SETDB1 KO cells, the initial cell diameter is set to $d_c = 21.8 \ \mu\text{m}$, with a corresponding nuclear diameter of $d_n = 10.9 \ \mu\text{m}$. Finally, to replicate the experimental setup described in \cite{Jebane:2023}, the microchannel width is fixed at $w = 6 \ \mu\text{m}$. The initial computational geometry adopted in the simulations is shown in Fig.~\ref{fig:setup}.

Second, the following numerical parameters were used in all simulations. We set the time step to $\tau = 2.5 \times 10^{-3} , \mathrm{s}$, and the characteristic length, the typical mesh element size, to $l_c = 5.0 \times 10^{-2} \, \mu\mathrm{m}$, the same value for both the cell plasma membrane and the nuclear envelope. Together, these choices guarantee stability and convergence of the numerical scheme while keeping an appropriate balance between accuracy and computational cost.

Determining an appropriate value for the viscous friction coefficient, $\omega$, proved challenging for both the cell membrane and the nuclear envelope due to the scarcity of direct measurements in the literature. Since this parameter strictly governs the temporal dynamics of the system, specifically the overall time required for the cell to totally enter the microchannel, we adopted the rationale of \cite{Larripa:2007} and calibrated it against the experimental data reported in \cite{Jebane:2023}. We fix $\omega_c = 0.5 \, \text{g}/(\mu\text{m}^2 \cdot \text{s})$ as a constant reference value for the cell plasma membrane; this specific choice carries no loss of generality, since a different value could always be absorbed into a rescaling of time. Conversely, the nuclear friction coefficient, $\omega_n$, is not treated as a constant. To capture the dynamic remodelling of the nucleus, $\omega_n$ is modelled as a function of the local value of the molecular state variable $W$ through the function $f_1(W)$.

Next, appropriate values must be assigned to the mechanical parameters governing the deformation of the cell and nuclear membranes, namely the surface tension and bending stiffness coefficients. These parameters play a fundamental role in determining cell morphology and the ability of the nucleus to deform while migrating through confined environments. Although baseline values can be informed by experimental measurements available in the literature \cite{Fischer:2014, Tinevez:2009, Salbreux:2012, Himbert:2022}, the present model requires distinct parameter sets to reproduce the experimentally observed mechanical differences between Ctrl and SETDB1 KO cells. As previously introduced, the nuclear surface tension and bending stiffness depend dynamically on the local value of the molecular state variable $W$ through the interpolation functions $f_2(W)$ and $f_3(W)$.

In particular, according to the constitutive relation \eqref{f2}, reference values for the nuclear surface tension must be specified for both phenotypes. We denoted by $\hat{k}^{\mathrm{s}}_n$ and $\tilde{k}^{\mathrm{s}}_n$ the reference nuclear surface tensions associated with the Ctrl and SETDB1 KO conditions, respectively. Guided by the relative differences in nuclear stiffness measured using optical tweezers in \cite{Comunian:2025}, these parameters are calibrated as $\hat{k}^{\mathrm{s}}_n = 0.16025 \,\mathrm{nN}/\mu\mathrm{m}$, $
\tilde{k}^{\mathrm{s}}_n = 0.21346 \,\mathrm{nN}/\mu\mathrm{m}$.
These values reproduce the experimentally observed increase in nuclear stiffness associated with SETDB1 KO. For the cell plasma membrane, a fixed
lower value is adopted. Since the plasma membrane is assumed to be mechanically more compliant than the nucleus, its surface tension coefficient is taken to be one order smaller than the reference nuclear surface tension in the control condition.

For the bending stiffness, we similarly refer to the constitutive relation \eqref{f3}. The bending rigidity of the cell plasma membrane is prescribed as $k^{\mathrm{b}}_c = 10^{-4}\,\mathrm{nN}\cdot\mu\mathrm{m}$. This value is chosen within the range of bending rigidities reported for biological membranes \cite{Himbert:2022}. Consistent with the higher effective mechanical rigidity of the nuclear envelope, we prescribe the corresponding reference nuclear bending rigidities as $\hat{k}^{\mathrm{b}}_n = 10^{-3}\,\mathrm{nN}\cdot\mu\mathrm{m}$ for the control condition and $\tilde{k}^{\mathrm{b}}_n = 10^{-2}\,\mathrm{nN}\cdot\mu\mathrm{m}$ for the SETDB1 KO condition. Thus, the reference nuclear bending rigidity is one order larger than that of the plasma membrane in the control condition, and two orders larger in the SETDB1 KO condition, consistently reflecting the increased nuclear stiffness associated with SETDB1 depletion.

The penalty parameters $\Lambda_A$ and $\Lambda_P$ in Eq.~\eqref{lambda_m} were tuned to enforce a reasonable constraint on the variations of the cell area and perimeter, respectively. Accordingly, we set $\Lambda_A = 1.0 \, \text{nN}/\mu\text{m}^4$ and $\Lambda_P = 0.1 \, \text{nN}/\mu\text{m}^2$, both for the cell and the nucleus.

All parameters governing the self-repulsive interaction were tuned to reproduce the anticipated qualitative behaviour: a self-interaction strength per unit area of $k_{\mathrm{sr}} = 1.0 \, \text{kPa}$, a decay sharpness $\kappa_d = 5.0 \, \mu\text{m}^{-1}$, and a distance threshold $\delta_d = 0.1 \, \mu\text{m}$. The sensitivity to the relative orientation between normals was fixed at $\kappa_n = 5.0$, with $\delta_n = -0.7$ as the corresponding orientation threshold.

Turning to the external force contributions, the barrier force per unit area was assigned an amplitude $k_{\mathrm{bar}} = 8.0 \, \text{kPa}$ and a steepness $\beta = 5.0 \, \mu\text{m}^{-1}$. The pushing pressure driving the cell, $k_{\mathrm{pr}}=16.0 \, \text{kPa}$, was taken directly from the experimental measurements in \cite{Jebane:2023}. Cell–nucleus repulsion was penalized with strength $k_{\text{rep}} = 18.0 \, \text{kPa}$ and steepness $\alpha = 5 \, \mu\text{m}^{-1}$, while a magnitude of $k_{\mathrm{el}} = 2.0 \, \text{kPa}$ was assigned to the elastic force keeping the cell plasma membrane anchored close to the nucleus.

\begin{table}[ht!]
\centering
\begin{tabular}{|c|c|c|}
\hline
\textbf{Par.} & \textbf{Value} & \textbf{Refs.} \\
\hline
$d_c$ (Ctrl) & 19.4 $\mu$m & \cite{Comunian:2025} \\
$d_n$ (Ctrl) & 9.7 $\mu$m & estimated \\
$d_c$ (SETDB1 KO) & 21.8 $\mu$m & \cite{Comunian:2025} \\
$d_n$ (SETDB1 KO) & 10.9 $\mu$m & estimated \\
$w$ & 6 $\mu$m & \cite{Jebane:2023} \\
$\tau$ & $2.5 \times 10^{-3}$ s & estimated \\
$l_c$ & $5.0 \times 10^{-2}$ $\mu$m & estimated \\
$\omega_c$ & 0.5 $\text{g}/\left[(\mu \text{m})^2 \cdot \text{s}\right]$ & estimated \\
$\hat{k}^{\mathrm{s}}_n$ (Ctrl) & 0.16025 nN/$\mu$m & \cite{Comunian:2025} \\
$\tilde{k}^{\mathrm{s}}_n$ (SETDB1 KO) & 0.21346 nN/$\mu$m & \cite{Comunian:2025} \\
$k^{\mathrm{s}}_c$ & 0.016025 nN/$\mu$m & estimated \\
$k^b_c$ & $10^{-4}$ nN$\cdot\mu$m & \cite{Himbert:2022} \\
$\hat{k}^b_n$ (Ctrl) & $10^{-3}$ nN$\cdot\mu$m & estimated \\
$\tilde{k}^b_n$ (SETDB1 KO) & $10^{-2}$ nN$\cdot\mu$m & estimated \\
$\Lambda_A$ & 1.0 nN$/\mu\text{m}^4$ & estimated \\
$\Lambda_P$ & 0.1 nN$/\mu\text{m}^2$ & estimated \\
$k_{\mathrm{sr}}$ & 1.0 kPa & estimated \\
$\kappa_d$ & 5.0 $\mu \mathrm{m}^{-1}$ & estimated \\
$\delta_d$ & 0.1 $\mu$m & estimated \\
$\kappa_n$ & 5.0 & estimated \\
$\delta_n$ & -0.7 & estimated \\
$k_{\mathrm{bar}}$ & 8 kPa & estimated \\
$\beta$ & 5.0 $\mu \mathrm{m}^{-1}$ & estimated \\
$k_{\mathrm{pr}}$ & 16 kPa & \cite{Jebane:2023} \\
$k_{\text{rep}}$ & 18 kPa & estimated \\
$\alpha$ & 5 $\mu \mathrm{m}^{-1}$ & estimated \\
$k_{\mathrm{el}}$ & 2 kPa & estimated \\
\hline
\end{tabular}
\caption{Reference parameters derived from biophysical experiments and literature}
\label{tab:parameters}
\end{table}
\section{Numerical simulations and results}
\label{sec:results}
In this section, we present the numerical simulations conducted to describe the proposed mathematical framework. The primary objective is to reproduce the macroscopic mechanical features of the two distinct biological scenarios described above: the Ctrl cancer cells and the SETDB1 KO cells, calibrating the model's behaviour against experimental observations. 

\subsection{Molecular dynamics at the nuclear envelope}
\label{sec:results_biochem}
Before analysing the macroscopic deformation of the cell, it is essential to characterize the temporal evolution of the molecular state variables, as the local value of $W$ at the nuclear periphery directly governs the mechanical properties of the nucleus. To this end, for simplicity and to isolate the purely kinetic behaviour, we neglect spatial diffusion and solve the associated zero-dimensional system of Ordinary Differential Equations (ODEs). The temporal evolution is thus entirely driven by the reaction terms $f(E,U,W)$, $g(E,U,W)$, and $h(E,U,W)$. 

The kinetic coefficients involved in the governing functions are set as follows:
$k_1 = 40.0 \, \text{s}^{-1}$, $k_2 = 250.0 \, \text{s}^{-1}$, $k_3 = 50.0 \, \text{s}^{-1}$, $k_4 = 25.0 \, \text{s}^{-1}$, $a = 10.0 \, \text{s}^{-1}$, $b = 50.0 \, \text{s}^{-1}$, and $c = 20.0 \, \text{s}^{-1}$.
These coefficients are selected as phenomenological parameters rather than being directly inferred from experimental data. Their values are chosen to reproduce the key qualitative features of the reaction network, namely the activation of downstream processes, the inhibitory effect of SETDB1 on SUV39H1, and the accumulation of H3K9me3 at the nuclear periphery.

\begin{figure}
    \centering
    \includegraphics[width=0.49\linewidth]{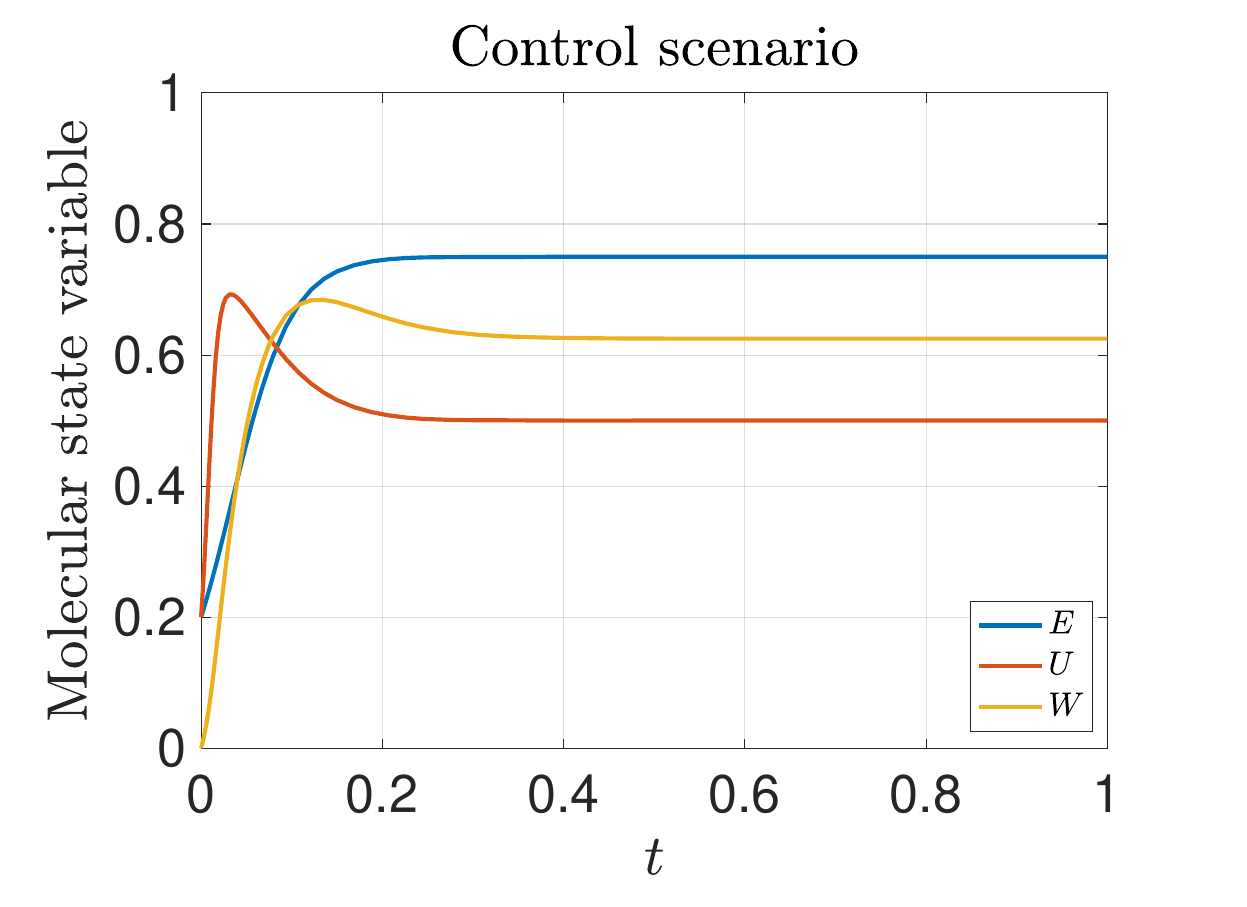} 
    \includegraphics[width=0.49\linewidth]{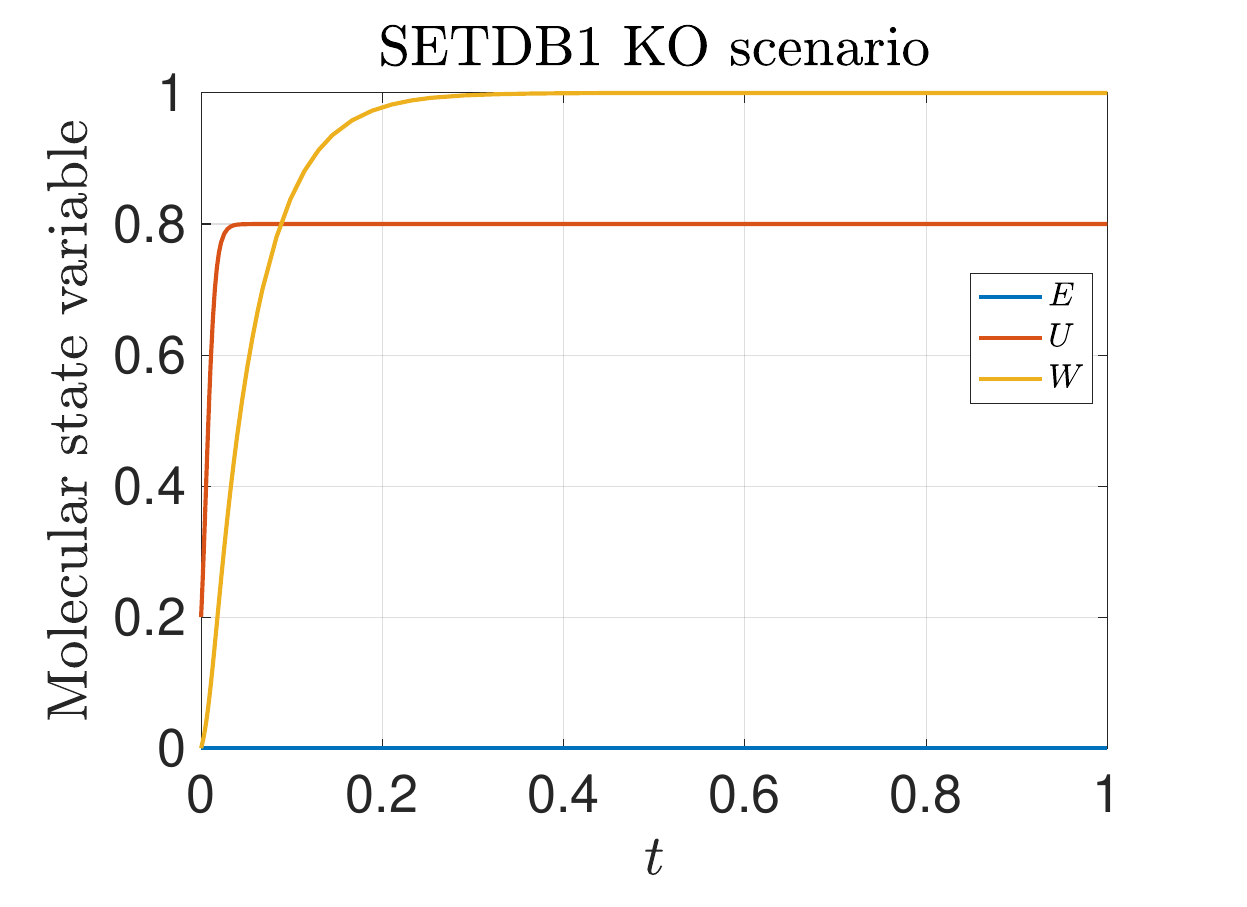}
    \caption{Molecular state variables $E$, $U$, and $W$ in the two studied scenarios.}
    \label{fig:ODE}
\end{figure}

For the control scenario, we initialize the ODE system using the spatially homogeneous condition introduced in Eq.~\eqref{IC_const}. The numerical integration demonstrates that the system transiently evolves and ultimately converges toward the stable asymptotic steady state $W_4^*$. Conversely, to simulate the mutant scenario where the expression of SETDB1 is biologically suppressed, we initialize the system with the following depleted state:
\begin{equation}
\label{IC_KO}
    (E, U, W)\big|_{t=0} = (0.0 \, , \, 0.2 \, , \, 0.0) \, . 
\end{equation}
Under this specific initial condition, the numerical results shown in Figure \ref{fig:ODE} demonstrate that the molecular system undergoes a distinct kinetic trajectory, inevitably reaching the alternative steady state $W_2^*$, which is higher than $W_4^*$. 

As mathematically formulated in Section~\ref{sec:functions}, the difference between these saturated states directly triggers the phenomenological divergence of the two cell lines. Specifically, the condition $W_4^* < W_2^*$ translates into an automated increase in the viscous friction, surface tension, and bending stiffness for the SETDB1 KO nucleus compared to the control one. Having validated the correct evolution of these homogeneous molecular states, the computational framework can now accurately feed the chemo-mechanical coupling to simulate the physical entry of the cells into the  microchannel.
\subsection{Microchannel entry dynamics and cellular deformation}

\begin{figure}
    \centering
    \includegraphics[width=\linewidth]{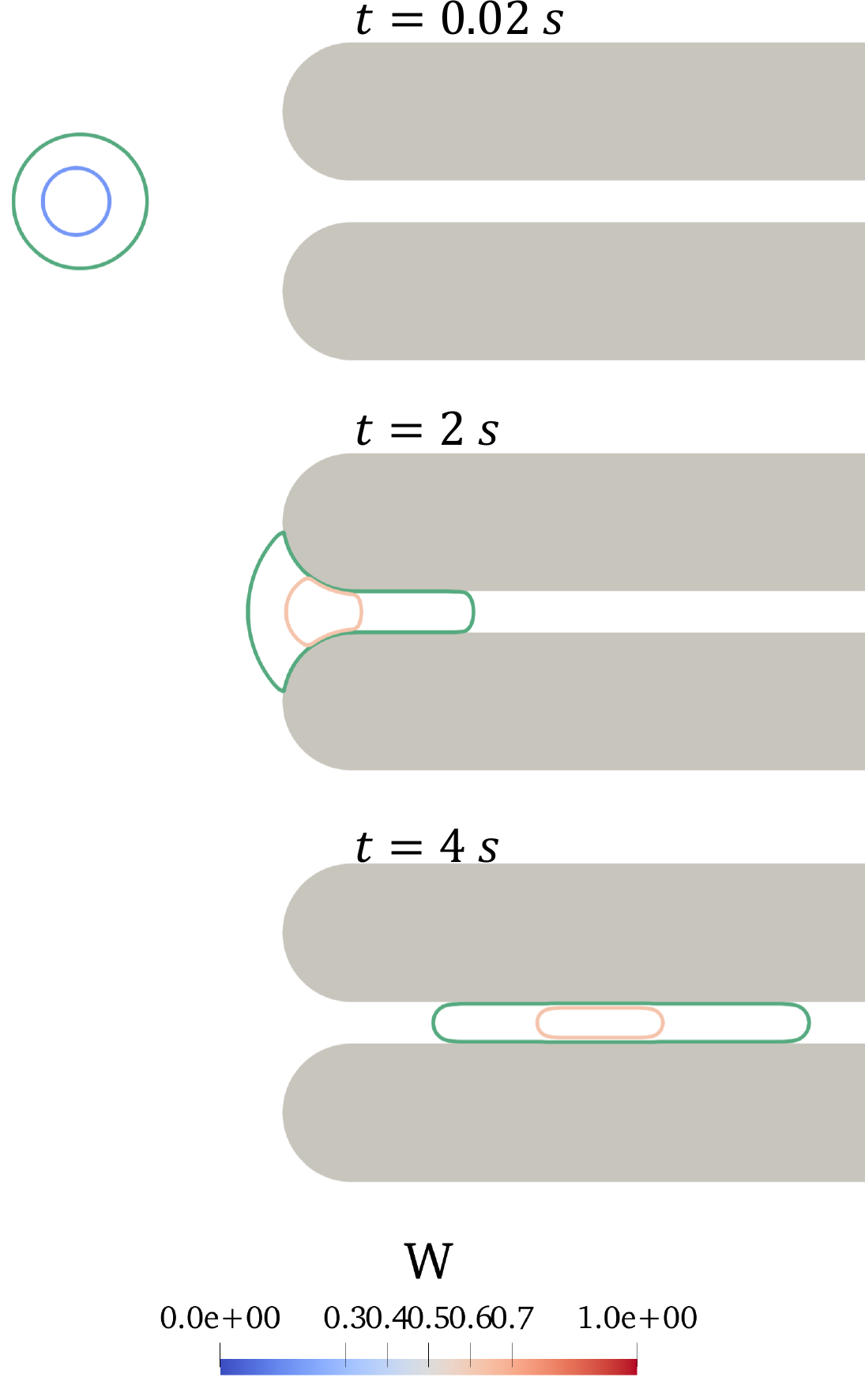}
    \caption{Snapshot of cell translocation through the microchannel. Control cell. Simulation performed using the initial condition \eqref{IC_const} for the molecular S-RDEs. The colour bar indicates the value of $W$, which is defined only on the nuclear envelope.}
    \label{fig:cell_Ctrl}
\end{figure}

\begin{figure}
    \centering
    \includegraphics[width=\linewidth]{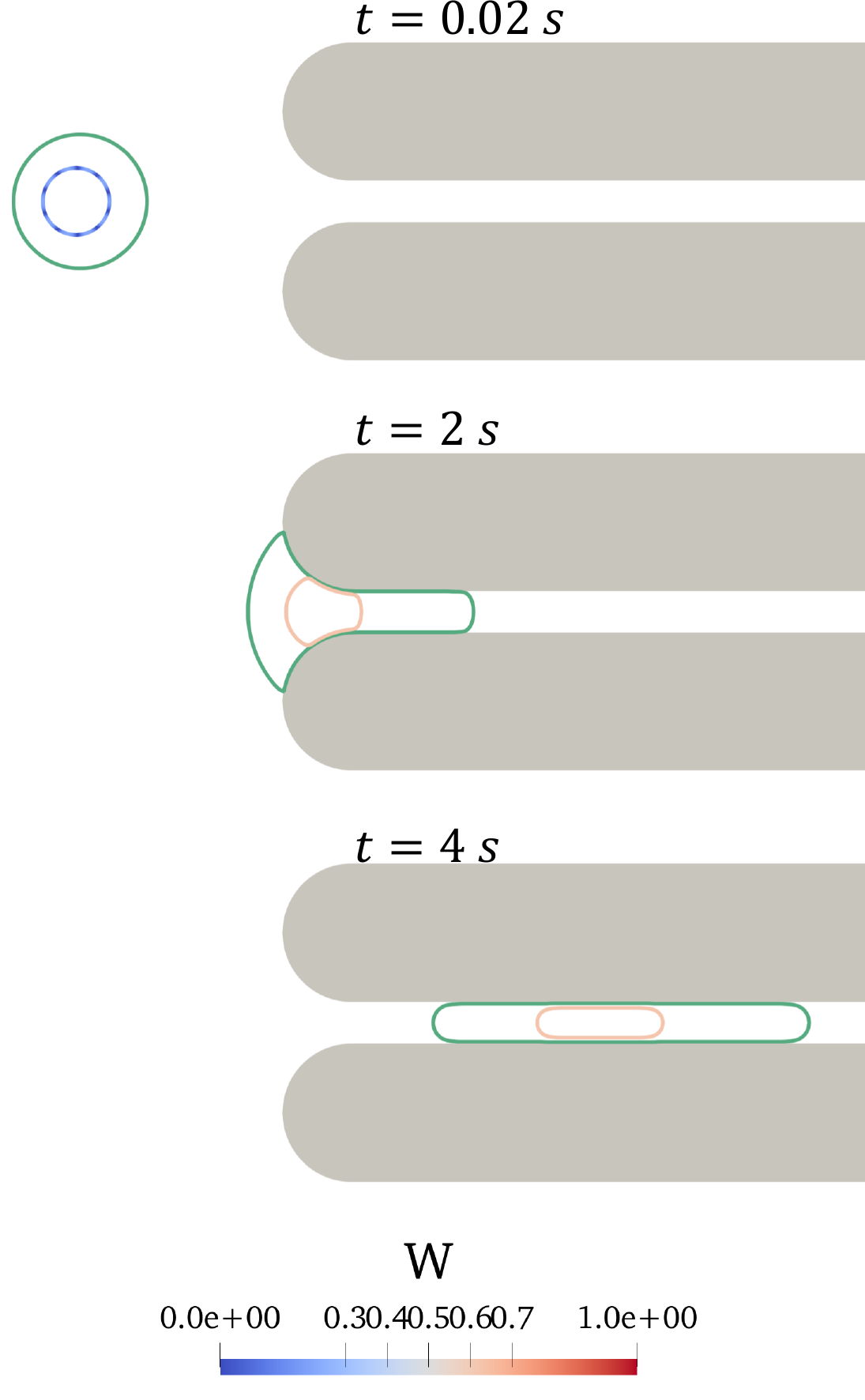}
    \caption{Snapshot of cell translocation through the microchannel. Control cell. Simulation performed using the spatially non-uniform initial condition \eqref{IC_cos} for the molecular S-RDEs. The colour bar indicates the value of $W$, which is defined only on the nuclear envelope.}
    \label{fig:cell_Ctrl_cos}
\end{figure}

We now present numerical simulations of the full model, using the spatial and temporal discretisation described in Section~\ref{sec:discretization}. The parameters used in the simulations are summarised in Table~\ref{tab:parameters} in Section~\ref{sec:results_biochem}. The diffusion coefficients are assumed to be identical for all molecular state variables, with $D_E=D_U=D_W=0.1\,\mu\mathrm{m}^2/\mathrm{s}$. We first consider Ctrl cells, for which two different initial conditions for the molecular state variables are prescribed, namely \eqref{IC_const} and \eqref{IC_cos}. The corresponding results are shown in Figure \ref{fig:cell_Ctrl} and Figure \ref{fig:cell_Ctrl_cos}, respectively.
In the first case, the molecular state variable $W$ evolves in time while remaining spatially homogeneous along the nuclear periphery $\Gamma_n$. In the second case, $W$ is initially distributed heterogeneously over $\Gamma_n$. However, this spatial heterogeneity is rapidly smoothed out, resulting in an approximately homogeneous distribution of $W$ along the nuclear periphery. Importantly, the characteristic timescale of this molecular redistribution is significantly shorter than the timescale associated with cell entry into the microchannel. Thus, the initial molecular spatial and temporal heterogeneity is almost entirely dissipated before the onset of cell translocation through the confined region. Following this short transient regime, both initial conditions converge to the same value of $W$, leading to essentially indistinguishable mechanical responses. Consequently, the cell exhibits the same translocation behaviour in both cases and successfully enters the microchannel, with an entry time of $2.17\,\mathrm{s}$.

\begin{figure}
    \centering
    \includegraphics[width=\linewidth]{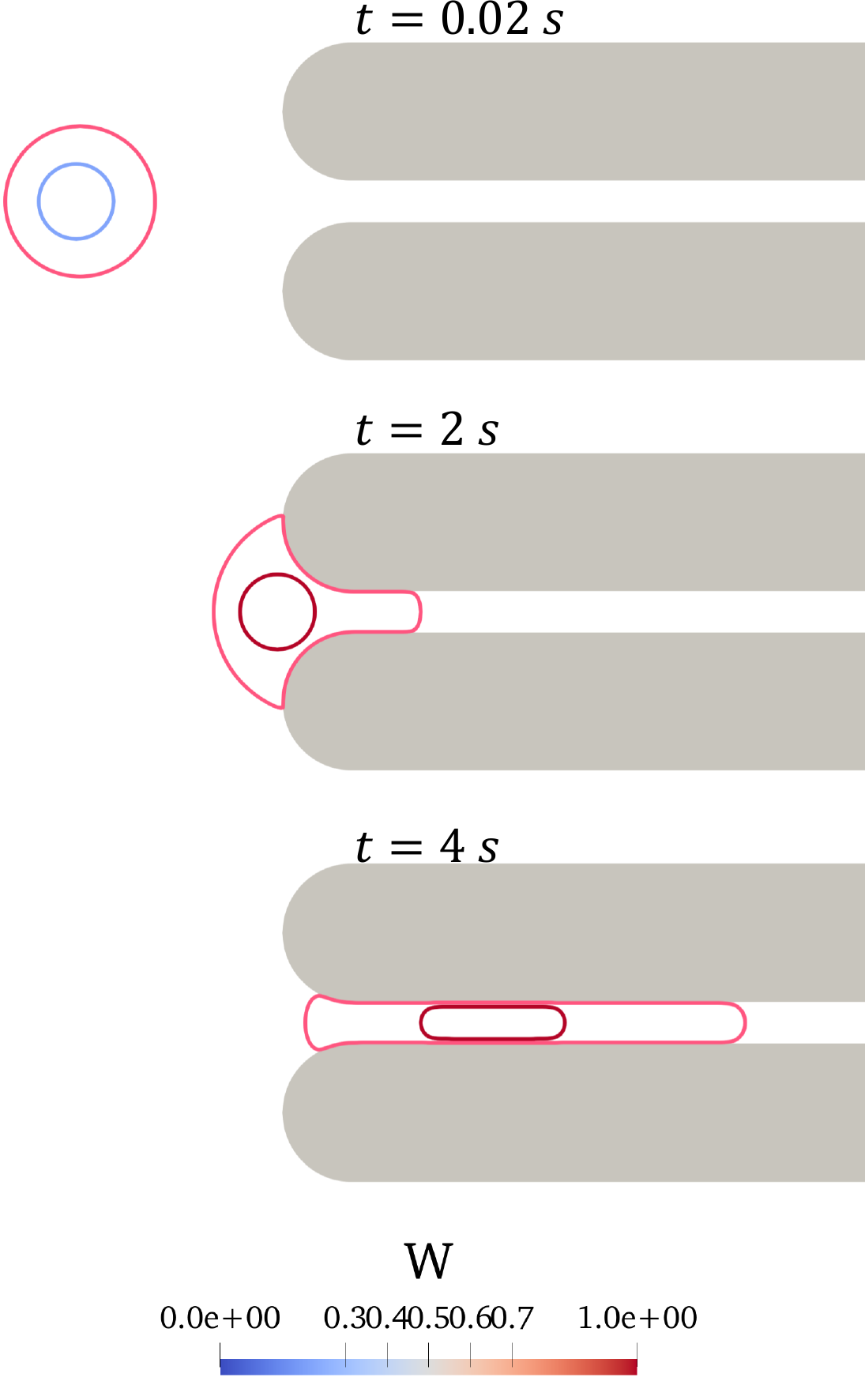}
    \caption{Snapshot of cell translocation through the microchannel. SETDB1 KO cell. Simulation performed using the initial condition \eqref{IC_KO} for the molecular S-RDEs. The colour bar indicates the value of $W$, which is defined only on the nuclear envelope.}
    \label{fig:cell_KO}
\end{figure}

We next consider the SETDB1 KO scenario, initialised with the condition \eqref{IC_KO}. In this case, the system converges to a different, higher homogeneous steady-state value of $W$ on $\Gamma_n$ before the cell enters the microchannel, reflecting the altered molecular regulation associated with SETDB1 depletion. This change in the molecular state modifies the mechanical properties of the nuclear envelope, resulting in an effectively stiffer nuclear response and a slower translocation through the confined channel. As shown in Figure \ref{fig:cell_KO}, cell migration is slightly delayed, with an entry time of $3.22\,\mathrm{s}$, compared with $2.17\,\mathrm{s}$ in the control case.
The complete time evolution of all the simulations is provided in Supplementary Videos S1 and S2.

\begin{figure}
    \centering
    \includegraphics[width=1\linewidth]{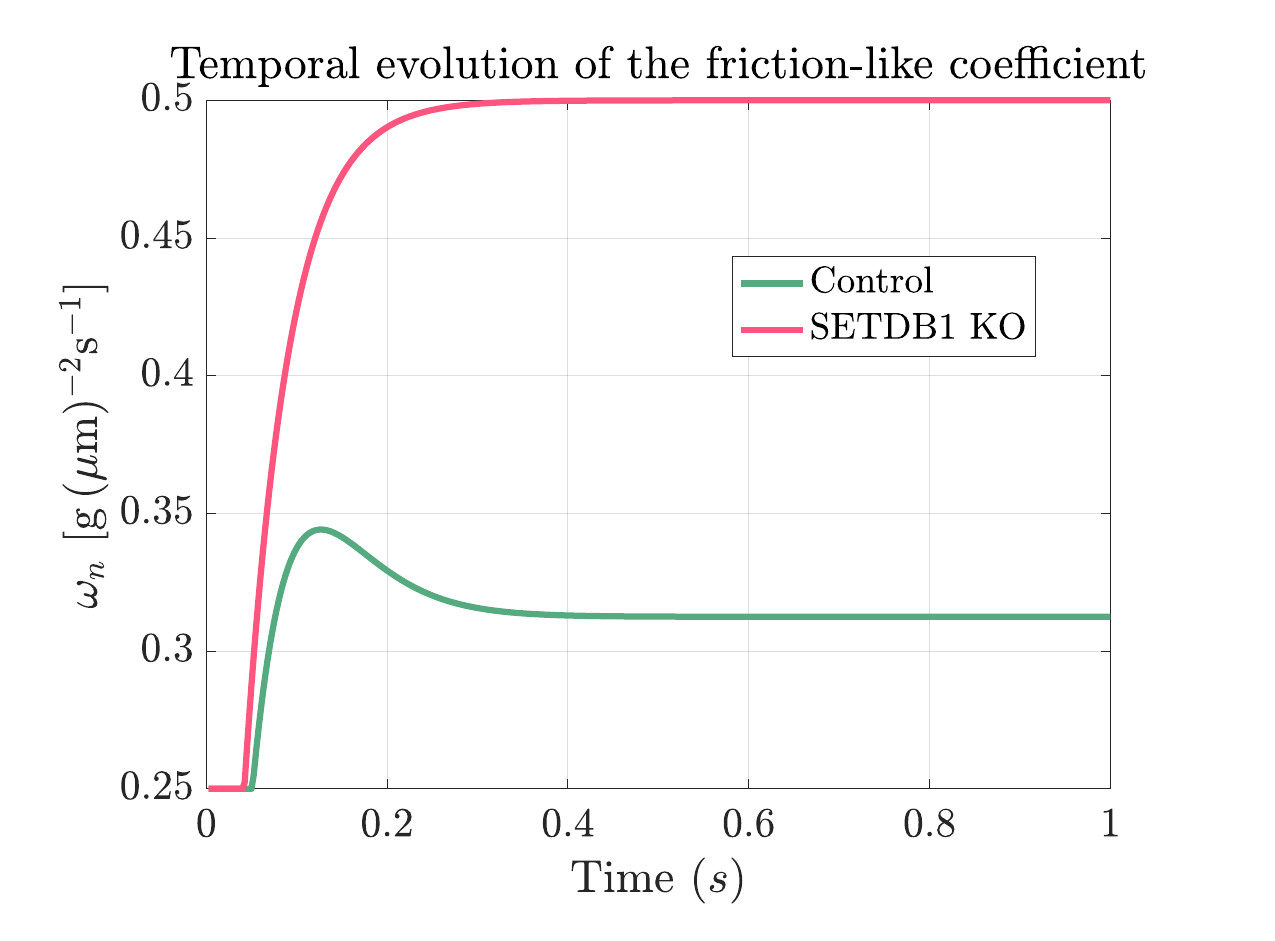}
    \includegraphics[width=1\linewidth]{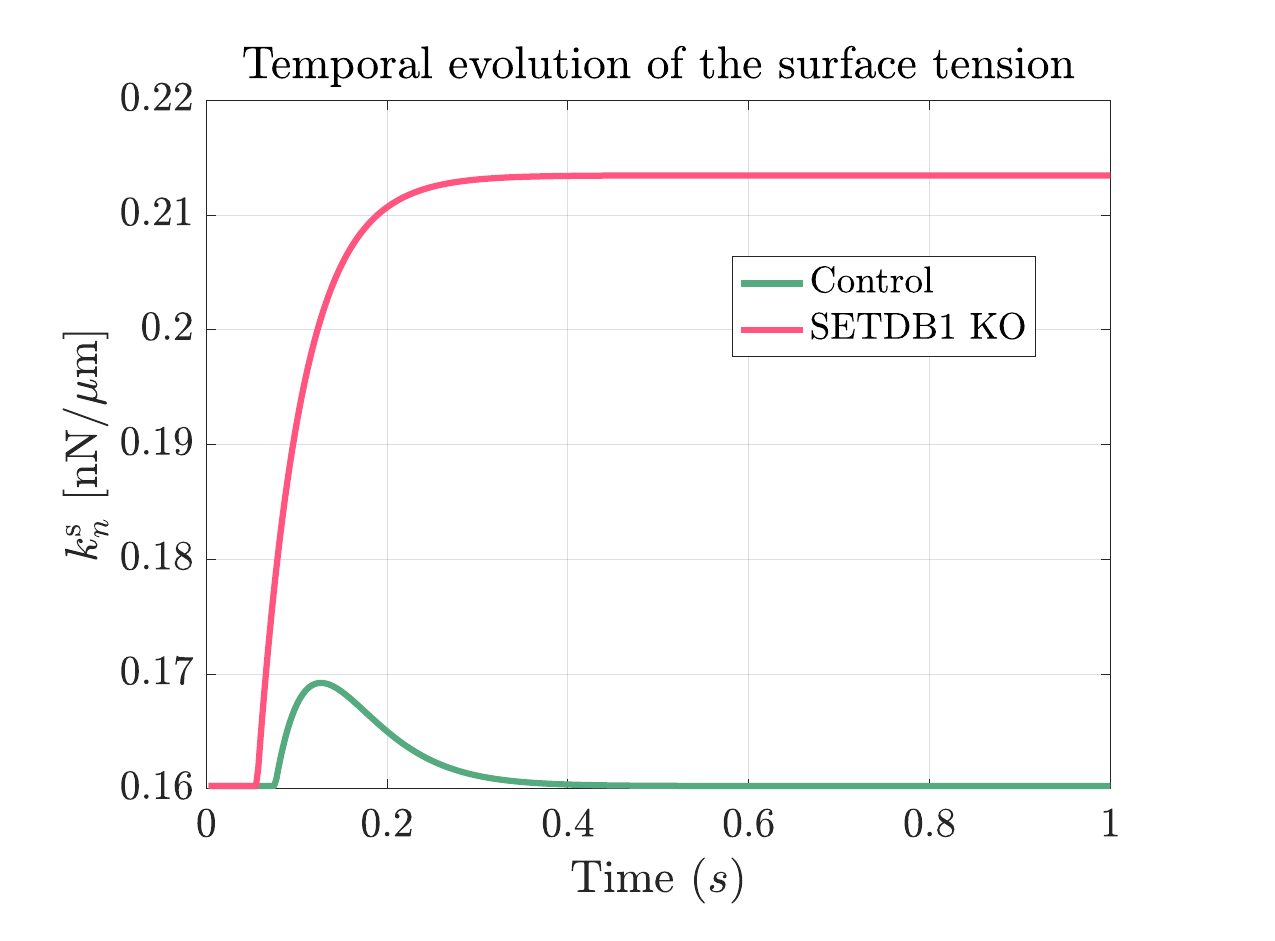}
    \includegraphics[width=1\linewidth]{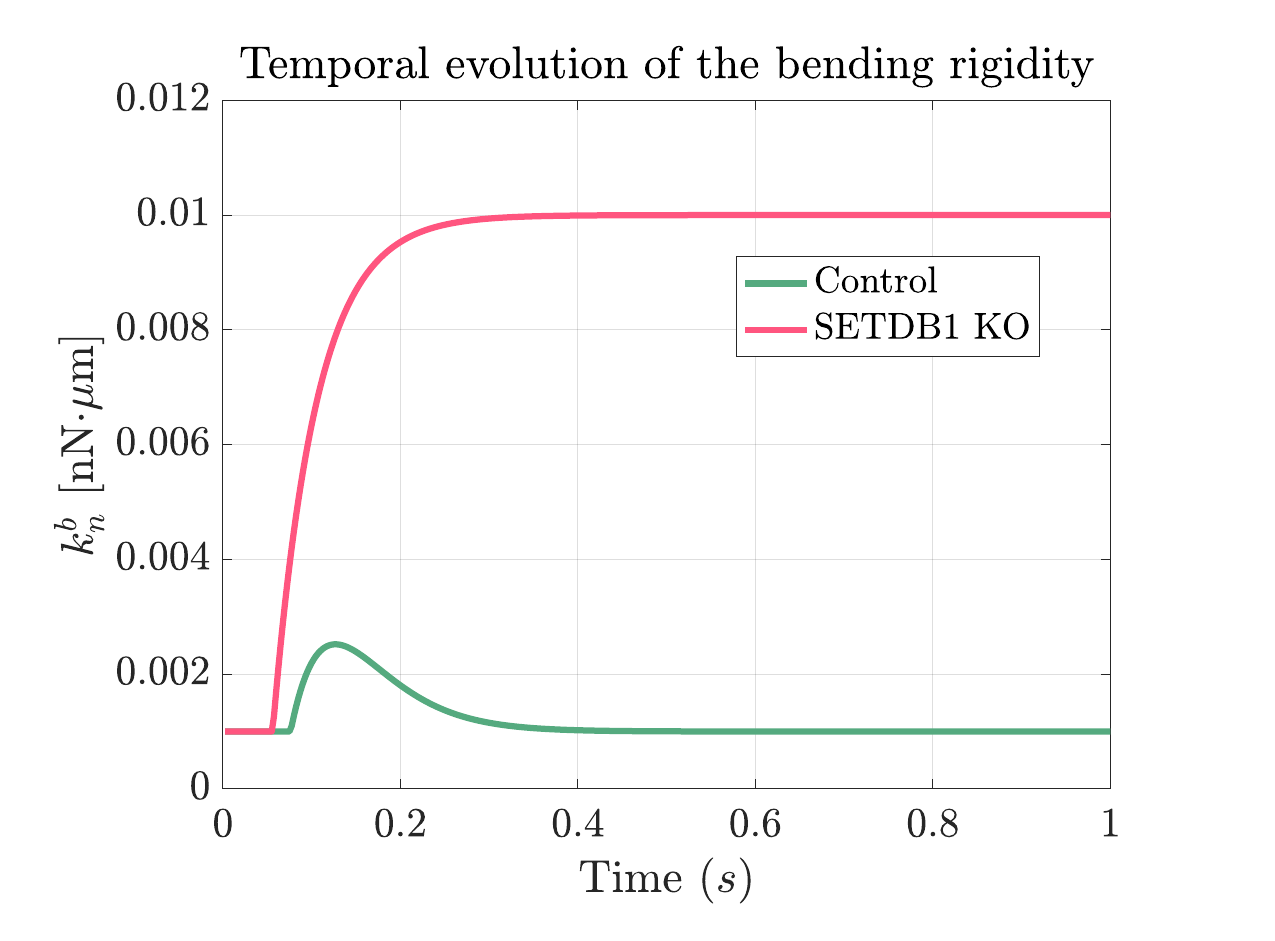}
    \caption{Temporal evolution of the averaged nuclear mechanical parameters for control and SETDB1 KO cells. The spatially averaged friction-like coefficient $\omega_n$, surface tension coefficient $k_n^{\mathrm{s}}$, and bending rigidity $k_n^{b}$ are reported during the initial phase preceding cell entry into the microchannel.}
\label{fig:mechanical_parameters_evolution}
\end{figure}

To further elucidate the coupling between chromatin regulation and the resulting nuclear mechanical response, we analyse the temporal evolution of the main mechanical parameters governing nuclear envelope dynamics. Specifically, we investigate the friction-like coefficient $\omega_n$, the surface tension coefficient $k_n^{\mathrm{s}}$, and the bending rigidity $k_n^{b}$ during confined migration for both control and SETDB1 KO conditions. Since these quantities are defined as spatially varying fields on the evolving nuclear envelope $\Gamma_n(t)$, we compute their spatial averages over the entire nuclear boundary at each time step. These averaged values provide a global characterisation of the mechanical state of the nucleus and allow us to quantify how different molecular states affect nuclear deformation and translocation through the confined microchannel. The temporal evolution of the averaged mechanical parameters for the control and SETDB1 KO cases is shown in Figure \ref{fig:mechanical_parameters_evolution}. The figure shows the temporal evolution of the averaged mechanical parameters from the beginning of the simulation up to $1\,\mathrm{s}$, corresponding to the initial phase preceding cell entry into the microchannel. During this interval, the mechanical parameters rapidly stabilise and reach their respective equilibrium values before the onset of significant nuclear deformation and confined migration. The comparison reveals how the altered chromatin organisation associated with SETDB1 depletion modifies the mechanical response of the nuclear envelope. In particular, the SETDB1 KO condition exhibits higher values of the averaged mechanical coefficients.

Overall, these simulations illustrate how a molecular state defined at the nuclear periphery can be coupled to nuclear mechanical properties within the proposed framework. The Ctrl and SETDB1 KO conditions provide a biological example to investigate how such nuclear mechanical changes contribute to confined migration.

An interesting aspect of the proposed modelling framework is that it provides a link between two experimentally accessible but distinct observables. On the one hand, optical tweezers provide a characterisation of nuclear mechanical properties, while, on the other hand, microfluidic experiments quantify the time required by the whole cell to enter a confined channel \cite{Comunian:2025,Jebane:2023}. In the present study, the experimentally observed difference in nuclear rigidity between Ctrl and SETDB1 KO cells, as measured using optical tweezers, is incorporated into the model through distinct baseline values of the nuclear envelope surface tension. These mechanical parameters are, in turn, dynamically modulated by the molecular state through the surface reaction--diffusion system. The resulting changes in nuclear mechanics affect nuclear deformation and the translocation dynamics of the whole cell. The proposed framework therefore provides a mechanistic link between nuclear-scale mechanical measurements and cell-scale migration behaviour.

In the simulations, the cell is considered to have fully entered the microchannel when the rear-most point of the cell crosses the left boundary of the channel. The entry time is therefore defined as the time elapsed between the initial contact of the cell with the channel and the instant at which the rear of the cell reaches the channel entrance. The experimentally measured average entry times are $1.44\,\mathrm{s}$ for Ctrl cells and $5.01\,\mathrm{s}$ for SETDB1 KO cells, whereas the corresponding simulations yield entry times of $2.17\,\mathrm{s}$ and $3.22\,\mathrm{s}$, respectively. Thus, increasing nuclear mechanical resistance reproduces the direction of the experimentally observed phenotype but accounts for only part of the difference between Ctrl and SETDB1 KO cells.
This discrepancy is consistent with the experimental characterization of these two cell populations. SETDB1 depletion affects not only nuclear mechanics, but also the mechanical properties of the whole cell \cite{Zakharova:2022}. By primarily varying the nuclear mechanical contribution, the present simulations therefore isolate one component of a multifactorial experimental phenotype. The moderate increase in simulated entry time suggests that changes in nuclear mechanics contribute to confined migration but are not sufficient, on their own, to reproduce the full experimental difference between Ctrl and KO cells. Additional changes in whole-cell mechanical parameters are required to obtain a larger separation between the two conditions.

More generally, these results illustrate the ability of the proposed framework to independently modulate mechanical parameters associated with different cellular compartments and to assess their respective contributions to confined migration. Here, the SETDB1/SUV39H1/H3K9me3 system provides a biologically characterized example in which a molecular state at the nuclear periphery is coupled to nuclear mechanics. The framework is therefore not restricted to this specific molecular pathway and could be adapted to other molecular processes occurring at cellular or nuclear interfaces.

\section{Conclusion and future perspectives}
\label{sec:conclusions}
In this work, we have presented a novel mathematical framework coupling geometric surface partial differential equations (GS-PDE) with surface reaction--diffusion equations (S-RDE) to investigate the interplay between molecular regulation and nuclear mechanics during confined cell migration. The cell plasma membrane and the nuclear envelope are described as evolving closed surfaces governed by geometric evolution laws accounting for the main mechanical properties involved in cell deformation, including surface tension, bending rigidity, and frictional resistance. The cell and its nucleus are subsequently simulated during translocation through a confined microchannel under an externally applied pressure difference.

The main novelty of the proposed framework lies in the integration of the mechanical description with a surface reaction--diffusion system posed on the evolving nuclear envelope. The framework is general and can be adapted to describe different molecular processes occurring at the nuclear periphery. In this specific work, the molecular component is used to model the dynamics of the SETDB1--SUV39H1--H3K9me3 pathway and its effects on the organisation of the tightly packed heterochromatin located beneath the nuclear envelope, with the reaction terms formulated following the experimental observations reported in \cite{Comunian:2025}. In particular, the model incorporates the experimentally identified effect of SETDB1 depletion in lung cancer cells, whereby the restoration of H3K9me3 enrichment at the nuclear periphery alters chromatin organisation and, consequently, affects the mechanical properties of the nucleus.

The numerical simulations demonstrate that the proposed model is able to reproduce the qualitative biological behaviour observed experimentally. Specifically, the altered molecular state associated with SETDB1 KO leads to changes in the nuclear mechanical parameters, resulting in a stiffer nuclear response and a slight increase in the entry time, suggesting that additional changes at the whole-cell level are required to reproduce the experimentally observed increase in the entry time. Furthermore, we investigated different initial molecular configurations, including spatially heterogeneous distributions, to account for distinct chromatin states and to assess the robustness of the mechanical response with respect to the initial molecular conditions.

Overall, the presented framework provides a first descriptive approach to study how molecular alterations at the nuclear envelope regulate nuclear mechanics and ultimately influence confined cell migration. Although the current model relies on a minimal phenomenological representation of the SETDB1--SUV39H1--H3K9me3 regulatory cascade, it establishes a flexible computational platform that can be progressively extended towards more complex biological scenarios.

Future developments will include the investigation of additional molecular and mechanical regulators involved in confined migration. In particular, extending the model to intermediate molecular perturbations, such as reduction of SETDB1 (=knockdown) instead of complete knockout, could provide a more direct assessment of the contribution of nuclear mechanics to cell translocation while limiting the broader cellular alterations associated with complete SETDB1 depletion. Comparing simulated control and knockdown conditions with the corresponding experimentally measured nuclear mechanical properties could therefore help clarify whether changes in nuclear stiffness alone are sufficient to significantly alter whole-cell translocation dynamics. Further extensions could consider the perturbation of SUV39H1, which directly regulates the accumulation of H3K9me3 at the nuclear periphery. Simulating different SUV39H1 levels would allow the reaction--diffusion component of the model to be used to describe changes in H3K9me3 distribution and to investigate their consequences for nuclear mechanics and cell translocation. Such perturbations would provide additional tests of the proposed mechanochemical coupling and could help disentangle the contributions of nuclear and cell-scale mechanical properties to confined migration.
From a mathematical perspective, it would be interesting to investigate the formulation of a coupled bulk--surface reaction--diffusion framework. In such a formulation, molecular dynamics would be described both within the nucleoplasm and on the evolving nuclear envelope. This extension would allow the representation of molecular processes involving soluble factors and transport mechanisms occurring throughout the nuclear bulk, while preserving the coupling between molecular regulation and nuclear mechanics. A further important direction concerns the introduction of a two-way mechano-chemical coupling. In the present model, molecular dynamics regulate the mechanical properties of the nuclear envelope, whereas the mechanical state does not influence the reaction--diffusion system. However, nuclear deformation, membrane tension, and changes in geometry may themselves affect molecular transport and reaction kinetics. Incorporating such feedback mechanisms would lead to a more comprehensive description of the reciprocal interplay between chromatin organisation, nuclear mechanics, and confined cell migration. Finally, extending the framework to three dimensions would open the possibility of investigating confinement scenarios beyond microchannel translocation, while requiring further developments in both the mathematical modelling and numerical implementation.

\backmatter

\bmhead{Supplementary information}
Two supplementary videos illustrating the simulated cell migration through the microchannel under the control and SETDB1 knockout conditions accompany this article.

\bmhead{Acknowledgements}
We thank the (EPI)2 Imaging platform (managed by Sandra Piquet and Audrey Chansard) - UMR7216 Epigenetic and Cell Fate Centre, for access to instruments and technical advice, with financial support from ITMO Cancer of Aviesan within the framework of the 2021-2030 Cancer Control Strategy, on funds administered by Inserm. We thank the GENIE platform, Epigenetic and Cell Fate center, for technical advice and practical help. We also thank Emmanuèle Helfer for performing the microfluidic experiments and providing the corresponding data, and Jean-Baptiste Manneville for providing access to the optical tweezers and for his technical support.

\bmhead{Declarations}

\begin{itemize}

\item \textbf{Funding:} This work was supported by the Agence National de la Recherche (ANR) through the MultiPhysC2M project (ANR-24-CE45-0690-03 grant to RA, SA, J.B. Manneville and E. Helfer) which also supports the postdoctoral fellowship of FB. This work was supported by the INSMI Maths Vives International Research Program (awarded to RA and AM), Visiting Professor Campaign, Université Côte d'Azur (awarded to RA and AM), the Canada Research Chair (Tier 1) in Theoretical and Computational Biology (CRC-2022-00147 to AM), the Natural Sciences and Engineering Research Council of Canada (NSERC), Discovery Grants Program (RGPIN-2023-05231 to AM), the British Columbia Knowledge Development Fund (awarded to AM), Canada Foundation for Innovation – John R. Evans Leaders Fund – Partnerships (awarded to AM), and the British Columbia Foundation for Non Animal Research (awarded to AM). FB is member of the National Group of Mathematical Physics (GNFM) of the Italian Institute for High Mathematics (INdAM). Work in the Ait-Si-Ali lab was supported by Fondation ARC (2025 PJA3 to SA), Ligue Nationale contre la Cancer - comité de Paris (to SA), Université Paris Cité through the Emergence call (ChromaStiff o SA). SC was supported by PhD fellowships from the doctoral school BioSPC of Paris Cité University and Fondation ARC pour la Recherche sur le Cancer.
All funders had no role in the study design, data collection and analysis, decision to publish, or preparation of the manuscript.

\item \textbf{Competing interests:} The authors declare no competing interests.

\item \textbf{Ethics approval and consent to participate:} Not applicable. 

\item \textbf{Consent for publication:} Not applicable.

\item \textbf{Data availability:} The datasets analysed during the current study are available from the corresponding author on reasonable request. 

\item \textbf{Materials availability:} Not applicable.

\item \textbf{Code availability:} The custom code developed for this study is open-source and available on GitHub at \url{https://github.com/francescaballatore/GS-PDEs_surface_RDEs}.

\item \textbf{Author contribution:} F.B. developed the mathematical model, implemented the computational code, performed the numerical simulations, and drafted the manuscript. S.C. provided the biological experimental data and contributed to the biological interpretation of the model and simulation results. A.M. supervised the research, conceptualized the study, contributed to the biological interpretation, and revised the manuscript. S.A. supervised the biological aspects of the study and contributed to the biological interpretation. R.A. supervised the research, conceptualized the study, contributed to the biological interpretation, and revised the manuscript. All authors read and approved the final manuscript.

\end{itemize}








\bibliography{bibliography}


\end{document}